# Orthorhombic and triclinic modifications of the arsenate $Cu_4O(AsO_4)_2$ and isotypic phosphates $Cu_4O(PO_4)_2$: strongly frustrated antiferromagnetics

**L M Volkova**

Institute of Chemistry, Far Eastern Branch, Russian Academy of Sciences, 690022 Vladivostok, Russia

E-mail: volkova@ich.dvo.ru

**Abstract**
A structural-magnetic models of the orthorhombic and triclinic modifications of the arsenates $Cu_4O(AsO_4)_2$ and isotypic phosphates $Cu_4O(PO_4)_2$ has been built and analyzed. Their base elements are the complicated ribbons composed of antiferromagnetic $Cu_4$ tetrahedra. Structurally, these tetrahedra have no shared copper atoms; however, there are strong antiferromagnetic (AFM) and ferromagnetic (FM) couplings between them, both within the complicated ribbons and between them. It has been established that both modifications are strongly frustrated 3-D antiferromagnetics due to competition between the nearest AFM interactions along the edges of the $Cu_4$ tetrahedra and competition between interactions and a multiplicity of long-range secondary AFM and FM interactions. Additionally, a large number of weaker long-range interactions are competing among each other. However, there is a possibility of the ordering Dzyaloshinskii-Moriya (DM) interaction in the centrosymmetric orthorhombic modification (*Pnma*), because two of the three types of magnetic ions, Cu1 and Cu3, are in the partial position 4c, where the ions are not related by the inversion center. In the triclinic modification (*P* -1) of $Cu_4O(As(P)O_4)_2$, all four copper ions are in the centrosymmetric equivalent position 2i, which prevents DM interactions. This centrosymmetry will allow magnetic interactions in the triclinic modification of $Cu_4O(As(P)O_4)_2$ to be still frustrated at lower temperature. It is possible that the triclinic modification of these compounds is a quantum spin liquid.

**Keywords**: structural-magnetic model, geometrical spin-frustrations, quantum spin liquid, Dzyaloshinskii-Moriya interaction, antiferromagnetic $Cu_4$ tetrahedron, volcanic mineral

## 1. Introduction

A large number of studies have been looking into the frustration of magnetic materials, because this matter is attractive both in terms of theoretical and experimental research [1-8]. Frustrated magnets are the materials in which localized magnetic moments (also known as spins) interact through competing exchange interactions that cannot, for geometric reasons, be satisfied at once. Due to frustration, states can be reached that are no long-range magnetic ordered ones. Special attention is being paid to the search for and research into frustrated magnetic compounds as these, because these may be the materials whose ground states are quantum spin liquids [4, 6]. Just as there is no such thing as a single type of a magnetic order, there is no such thing as a single type of quantum spin liquid either.

Different types of quantum spin liquid correspond to different models of long-range entanglement.

The main objective of this work was to explain the role of a perfectly ordered crystal structure in the emergence of the magnetic disorder known as *geometrical frustration*. For the purpose of our study, we will use only the simplest geometrical part of the huge body of scientific material devoted to frustrated magnets. This portion will be exemplified with three randomly interacting magnetic ions, which reside on two geometrical units, a triangle and a linear chain (figure 1) [1, 3, 4]. These geometrical units represent the elementary components of the crystal structures of magnetic compounds. Within the triangles and along the linear chains, frustration can exist at certain ratios of the strengths of magnetic interactions: either if $J_{12}$, $J_{13}$ and $J_{23}$ are antiferromagnetic at once ($J_{ij} < 0$) (figures 1(a) and (e)) or if one of them, for example, $J_{23}$, is anti-ferromagnetic ($J_{23} < 0$), while the other two, $J_{12}$ and $J_{13}$, are ferromagnetic ($J_{ij} > 0$) (figures 1(b) and (f)). By contrast, when $J_{12}$, $J_{13}$ and $J_{23}$ are ferromagnetic at once ($J_{ij} > 0$) (figures 1(c) and (d)) – or when only one of them, for example, $J_{23}$, is ferromagnetic ($J_{23} > 0$), and the other two, $J_{12}$ and $J_{13}$, are antiferromagnetic ($J_{ij} < 0$) (figures 1(g) and (h)), the system is not frustrated. The same phenomenon of ground state degeneracy – frustration – takes place in any closed spin chain consisting of an arbitrary number of spins if the product of the spin-spin interactions along the chain is negative.

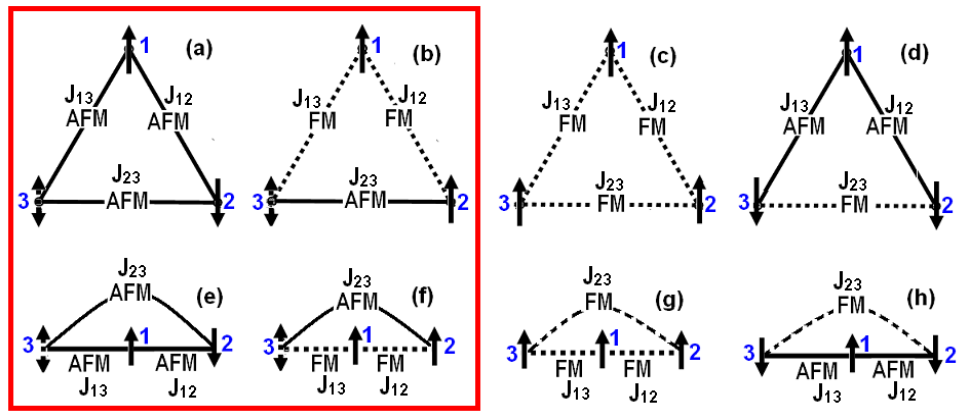

**Figure 1.** Frustration in a three-spin system. The system is frustrated when $J_{12}$, $J_{13}$ and $J_{23}$ are AFM ((a) and (e)) – or when $J_{23}$ is AFM, and the other two, $J_{12}$ and $J_{13}$, are FM ((b) and (f)). The system is not frustrated when $J_{12}$, $J_{13}$ and $J_{23}$ are FM ((c) and (g)) – or when $J_{23}$ is FM, and the other two, $J_{12}$ and $J_{13}$, are AFM ((d) and (h)).

However, the situation becomes much more complicated when it comes to real systems. A huge number of spin-spin interactions $J_{ij}$, which may have either sign and diverse strengths on the triangles, linear units and closed multi-spinned chains, can be observed in low-symmetry magnetic systems. This will produce a large number of frustrations. Dotsenko [1] showed that, to describe the state of complex systems come, the concept of self-averaging should be introduced even if the system of spin-spin interactions $J_{ij}$ being considered is fixed.

The purposes of our research were (1) to find complex strongly frustrated low-symmetry magnetic compounds that have a simple chemical composition and formula and (2) to build their structural-magnetic models. Our previous efforts [9, 10] suggest that the framework for the geometrical frustration of magnetic systems to happen could be oxocentered $OCu_4$ tetrahedra, which are basic to the crystal structures of most minerals at the Tolbachik volcano, the Kamchatka Peninsula [11, 12], and so we decided to take these minerals to our study. Pekov et al. [12] showed that the diversity and originality of fumarole systems of oxidizing type in this volcano are mineralogically unique. From a huge number of structural material [11, 12], we chose, for our purposes, two polymorphic modifications (that is, different crystal forms of the same chemical compound) of the arsenate $Cu_4O(AsO_4)_2$ [13]: triclinic ericlaxmanite and orthorhombic kozyrevskite – and the modifications of the phosphate $Cu_4O(PO_4)_2$ isostructural (isotypic) to them [14, 15]. These compounds consist of two types of centered tetrahedra: $OCu_4$ and $As(P)O_4$, where oxygen ions play a dual role. In $OCu_4$, the oxygen ion occupies the center of the tetrahedron, while in $As(P)O_4$, its corners. The oxygen ion is located centrally in the tetrahedron made of magnetic ions and plays a pivotal role in defining the magnetic properties of the material.

We will (1) calculate the parameters of the spin-spin interactions $J_{ij}$ in these four magnetic materials using the Crystal Chemistry Method [16 - 18]; (2) construct their structural-magnetic models; and (3) demonstrate a strong frustration of the magnetic system of these compounds, to raise awareness of these materials among theorists and experimenters. The structural-magnetic models makes it possible to reveal main correlations between the structures and magnetic properties of the compounds and thus to determine the crystal chemistry criteria for a targeted search for new functional magnetics.

## 2. Method of calculation

The structural-magnetic models are based on crystal chemical parameters (crystal structure, ion charge and ion size). The characteristics of these models include: (1) the sign and strength of magnetic interactions $J_{ij}$; (2) the dimensionality of magnetic structures (this does not always coincide with the dimensionality of the crystal structures); (3) the presence of magnetic frustrations in specific geometric configurations; (4) a possibility to reorient magnetic moments (that is, to enable AFM-to-FM transitions) due to displacement of intermediate ions located at critical positions.

To infer the sign (type) and strength of the magnetic interactions $J_{ij}$ from structural data, we used the Crystal Chemistry Method, our previous development, and the associated software program MagInter [16-18]. The Crystal Chemistry Method puts together three well-known concepts about the nature of magnetic interactions: Kramers's idea [19], the Goodenough–Kanamori–Anderson's model [20-22] and the polar Shubin–Vonsovsky's model [23].

The crystal chemistry method enables one to determine the sign (type) and strength of magnetic interactions $Jij$ on the basis of structural data. Within the framework of our consideration, the parameter $Jij$ and the exchange integral $Jij$ are practically synonymous - these quantities differ only by the scaling factor K, so we will use both terms interchangeably. According to this method, the coupling between magnetic ions $M_i$ and $M_j$ emerges in the moment of crossing the boundary between them by an intermediate ion ($A_n$) with the overlapping value of ~0.1 Å (figure 2). In the $Cu_4O(As(P)O_4)_2$ compounds under consideration, the magnetic ions are copper ions $Cu^{2+}$. The area of the limited space (local space) between the $M_i$ and $M_j$ ions along the bond line is defined as a cylinder, whose radius is equal to these ions radii. The strength of magnetic couplings and the type of magnetic moments ordering in insulators are determined mainly by the geometrical position and the size of intermediate ions $A_n$ in the local space between two magnetic ions ($M_i$ and $M_j$).

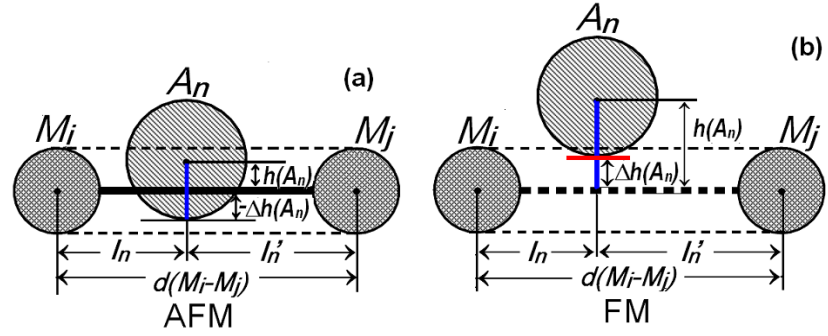

**Figure 2.** A schematic representation of the intermediate $A_n$ ion arrangement in the local space between magnetic ions $Mi$ and $Mj$ in cases where the $An$ ion initiates the emerging of the antiferromagnetic (**a**) and ferromagnetic (**b**) interactions. $h(A_n)$, $l_n$, $l_n$', and $d(M_i–M_j)$ are the parameters determining the sign and strength of magnetic interactions $(J_n)$.

The positions of intermediate ions $(A_n)$ in the local space are determined by the distance $h(A_n)$ from the center of the ion $A_n$ up to the bond line $M_i$-$M_j$ and the degree of the ion displacement to one of the magnetic ions expressed as a ratio $(l_n'/l_n)$ of the lengths $l_n$ and $l_n'$ ($l_n \leq l_n'$; $l_n' = d(M_i - M_j) - l_n$) produced by the bond line $M_i$-$M_j$ division by a perpendicular made from the ion center (figure 2).

The intermediate $A_n$ ions will tend to orient magnetic moments of $M_i$ and $M_j$ ions and make their contributions $j_n$ into the emergence of AFM or FM components of the magnetic interaction in dependence on the degree of overlapping of the local space between magnetic ions ($\Delta h(A_n)$), the asymmetry ($l_n'/l_n$) of position relatively to the middle of the $M_i$-$M_j$ bond line, and the distance between magnetic ions ($M_i$-$M_j$). Among the above parameters, only the degree of space overlapping between the magnetic ions $M_i$ and $M_j$ ($\Delta h(A_n) = h(A_n) - r_{A_n}$) equal to the difference between the distance $h(A_n)$ from the center of $A_n$ ion up to the bond line $M_i$-$M_j$ and the radius ($r_{A_n}$) of the $A_n$ ion determined the sign of magnetic interaction. If $\Delta h(A_n) < 0$, the $A_n$ ion overlaps (by $|\Delta h|$) the bond line $M_i$-$M_j$ and initiates the emerging contribution into the AFM-component of magnetic interaction. If $\Delta h(A_n) > 0$, there remains a gap (the gap width $\Delta h$) between the bond line and the $A_n$ ion, and this ion initiates a contribution to the FM-component of magnetic interaction.

The value of the contributions $j_n$ is defined by expressions:

$$j_n = \frac{\Delta h(A_n)\dfrac{l_n}{l_n'} + \Delta h(A_n)\dfrac{l_n'}{l_n}}{d(M_i - M_j)^2} \quad \text{(if } l_n'/l_n < 2.0\text{),} \quad (1)$$

and

$$j_n = \frac{\Delta h(A_n) \dfrac{l_n}{l_n'}}{d(M_i - M_j)^2} \quad \text{(if } l'_n / l_n \geq 2.0\text{)}. \tag{2}$$

The sign and strength of the magnetic coupling $J_{ij}$ are determined by the sum of the above contributions:

$$J_{ij} = \sum_n j_n \tag{3}$$

The $J_{ij}$ value is expressed in per angstrom units (Å$^{-1}$). If $J_{ij} < 0$, the type of magnetic moment ordering of $M_i$ and $M_j$ ions is antiferromagnetic, while if $J_{ij} > 0$, the type of magnetic moment ordering is ferromagnetic, if $J_{ij} = 0$ transition to the paramagnetic state.

It is possible to establish the reasons of occurrence of anomalies of magnetic interactions and magnetic phase transitions in magnets with the help of Eqs. (1) - (3). There exist several critical positions of intermediate $A_n$ ions when even a slight deviation from them could result in reorientation of magnetic moments (AFM–FM transition) and/or dramatic change of the magnetic interaction strength. It appears important to note that, under the effects of temperature, pressure, magnetic field, etc, the ions in a crystal structure could undergo displacement. That is why during prediction of possible changes in the sign and strength of magnetic interactions one should take into account not only the ions located exactly at critical positions, but also those in adjacent areas. The following intermediate ion positions can be considered as critical:

(a) $h(A_n) = r_M + r_{An}$: the distance $h(A_n)$ from the $A_n$ ion center to the bond line $M_i$–$M_j$ is equal to the sum of the $M$ and $A_n$ ionic radii. The $A_n$ ion reaches the surface of a cylinder of radius $r_M$, limiting the space area between the magnetic ions $M_i$ and $M_j$. In this case the $A_n$ ion does not induce the emerging of a magnetic interaction. However, on a slight decrease of $h(A_n)$ (the $A_n$ ion displacement inside this area) there emerges a strong FM interaction between magnetic ions.

(b) $h(A_n) = rA_n$ ($h(A_n) = 0$): the distance $h(A_n)$ from the center of the $An$ ion to the bond line $M_i$–$M_j$ is equal to the $A_n$ ionic radius (the $A_n$ reaches the bond line $M_i$–$M_j$). In this case the interaction between magnetic fields disappears. However, on a slight decrease of $h(A_n)$ (overlapping of the bond line by the $A_n$ ion) there emerges a weak AFM interaction, while on a slight increase of $h(A_n)$ (formation of a gap between the $A_n$ ion and the bond line $M_i$–$M_j$) there emerges a weak FM interaction.

(c) $l_n'/l_n = 2$: the $A_n$ ion is located at the boundaries of the central one-third of the space between magnetic fields. In this case the insignificant displacement of the $A_n$ ion to the center in parallel to the bond line $M_i$–$M_j$ results in a dramatic increase of the magnetic interaction strength.

In the case when there are several intermediate $A_n$ ions between the magnetic ions $Mi$ and $Mj$, the following critical positions are possible:

(d) When the ratio between the sums of the $j_n$ contributions to the AF and FM components of the interaction becomes close to 1, the interaction between the magnetic ions $M_i$ and $M_j$ is weak, and a slight displacement of even one of the intermediate $A_n$ ions could result in its complete disappearance or the AF–FM transition.

(e) When even one of the intermediate $A_n$ ions is in a critical position of (a) or (c) type, the contribution to AFM or FM components of the interaction could undergo dramatic changes because of even a slight displacement of these ions and, therefore, cause changes of respective scale in the interaction strength and reorientation of magnetic ion spins.

Calculation of magnetic moments in Bohr magnetons based on crystal chemical parameters was not provided. The crystal chemical method for calculating the parameters of magnetic interactions between magnetic ions is applicable to the study of both collinear and non-collinear magnets. We used this method to study 1-D and 2-D frustrated antiferromagnets, chiral magnetic solitons, potential spin liquids and multiferroics. This method was created by us to search for and predict new promising magnetic materials.

Thus, we have shown that that the structural–magnetic models of compounds built on the basis of calculations of magnetic couplings parameters by the crystal chemistry method make it possible to reveal the main correlations between the crystal structure of compounds and their magnetic properties. As a rule, under the temperature or pressure, structural phase transitions of magnetic compounds and displacement of intermediate ions located in critical positions are accompanied by magnetic transitions FM - AFM, FM - PM or AFM - PM. In this case, the structural transition temperature will also be the magnetic transition temperature.

We emphasize that the considered orthorhombic and triclinic modifications of the $Cu_4O(AsO_4)_2$ arsenate and isotypic $Cu_4O(PO_4)_2$ phosphate are polymorphs (i.e., different crystalline forms of the same chemical compound) and not the result of phase transitions.

We will consider pair exchange interactions, $J_{ij}$, not only between the nearest neighbors in the lattice, in the nodes of which they reside, but also at long distances. However, it should be noted that Crystal Chemistry Method overestimates the strength of interactions between magnetic ions at long distances ($d(M_i\text{-}M_j) \sim 8$ Å). Apparently, as the distance between magnetic ions increases, the rates of reduction in the strength of magnetic interaction become higher – and it becomes inversely proportional not to the square, but to the cube of the distance between them.

It should be particularly emphasized that the magnetics in question – the arsenate $Cu_4O(AsO_4)_2$ and the phosphate $Cu_4O(PO_4)_2$ – belong to a specific class of compounds, whose magnetic structure and properties are largely defined by two factors: the presence of Jahn-Teller (JT) $Cu^{2+}$ ions with orbital degeneracy [24-30] and the geometrical frustration of the magnetic couplings, both within the $Cu_4$ tetrahedra and between them. According to a large body of literature data and our crystal chemical studies [9, 31-32], intermediate X ions, whose bond with copper is JT elongated, do not contribute to magnetic coupling. Therefore, when calculating the parameters of the magnetic couplings $J$n using the Crystal Chemistry Method, we will neglect the contribution $j(X^{ax})$ made by the intermediate ions X at elongated positions to magnetic coupling with at least one of the two involved $Cu^{2+}$ ions.

Finally, it should be noted that, unlike any experimental setting, the Crystal Chemistry Method can calculate the ideal values of the magnetic parameters of separate couplings. This method ignores the potential impact of structural/"nonstructural" interactions and strengths on these couplings nor does it take account of the competition that weakens the couplings $J$n. Our calculations apply only to the regular lattices of the magnetic moments and intermediate ions that contribute to magnetic interactions.

To build the structural-magnetic models of the mineral polymorphs kozyrevskite $Cu_4O(AsO_4)_2$ (*Pnma*, ICSD-239833, [13]) and ericlaxmanite (*P* -l, ICSD-404850), [13]), we used their perfect synthetic analogs. The crystal structure of the synthetic mineral kozyrevskite (ICSD-81295 [33, 34]) was determined much more accurately (R value = 0.038) than that of its natural counterpart (R value = 0.1049). Ericlaxmanite was chosen for another reason. Its synthetic analog (ICSD-404850 [35]) is a stoichiometrically perfect crystal, while the Cu$^2$ position in its natural sample split into two subsites, each having an occupancy factor less than 100%.

The format of input data for the MagInter software program (crystallographic parameters, atom coordinates) is compatible with the cif-file in the Inorganic Crystal Structure Database (ICSD) (FIZ Karlsruhe, Germany). The ionic radii of Shannon [36] ($r(^{V}Cu^{2+})$ = 0.65 Å, $r(^{VI}O^{2-})$ = 1.40 Å, $r(^{IV}As^{5+})$ = 0.335 Å, $r(^{IV}P^{5+})$ = 0.170 Å) were used for calculations.

Tables 1 and 2 (Supplementary Note 1) show the crystallographic characteristics and parameters of magnetic couplings ($J$n) calculated on the basis of structural data and respective distances between magnetic ions in the materials under study. Additionally, the degree of overlapping of the local spaces between magnetic ions ($\Delta h(X)$), asymmetry ($l'_n/l_n$) of the position relative to the middle of the $Cu_i$–$Cu_j$ bond line, and the $Cu_i$–X–$Cu_j$ angle are presented for the intermediate ions X, which provide the maximal contributions ($j(X)$) to the AFM or FM components of these couplings $J$n. To translate the $J$n value in per angstrom (Å−1) into energy units more conventional for experimenters—millielectronvolt (meV)—one can use the scaling factor K = 74 ($J$n (meV) = 74 $J$n (A$^{-1}$)) [31].

## 3. Results and discussion

### *3.1. Coordination polyhedra and the crystal structure of a sublattice of magnetic ions $Cu^{2+}$*

The kozyrevskite $Cu_4O(AsO_4)_2$ (ICSD-81295) [33] (Supplementary Note 1, Table 1) crystallizes in the centrosymmetric orthorhombic space group *Pnma* (N62). The magnetic $Cu^{2+}$ ions occupy three crystallographically independent sites—Cu1, Cu2, and Cu3—and have a characteristic distortion of the $Cu^{2+}$ coordination polyhedra ($Cu1O_5$, a trigonal bipyramid, where d(Cu1-5O) = 1.842-2.109 Å; $Cu2O_5$, a distorted tetragonal pyramid, where d(Cu2-5O) = 1.912-2.337 Å; and $Cu3O_5$, a trigonal bipyramid, where d(Cu3-5O) = 1.920-2.182 Å) due to the Jahn-Teller effect enhanced by geometric hindrances related to the packing features (figure 3). Table 1 in Supplementary Note 1 also contains the crystallographic characteristics and parameters of the magnetic couplings, $J$n, of the orthorhombic phosphate $Cu_4O(PO_4)_2$ (ICSD-50459), with Cu2 replaced by Cu3, according to the original work [15]. The corresponding changes in the designations of the oxygen ions were taken into account, too.

The ericlaxmanite $Cu_4O(AsO_4)_2$ (ICSD-404850) [35] (Supplementary Note 1, Table 2), which crystallizes with centrosymmetric triclinic symmetry (space group *P* –1), has four main crystallographically independent Cu sites. The Cu(1) ions center distorted tetragonal pyramids (Cu(1)-4O = 1.926-1.959 Å of and Cu(1)-O8 = 2.646 Å), while the Cu(3) ions center distorted trigonal bipyramids (Cu(3)-4O = 1.911-2.015 Å and Cu(3)-O7 = 2.464 Å). The Cu(4) ions take positions in the elongated octahedra (Cu(4)-4O = 1.909-2.011 Å, Cu(4)-O(6) = 2.457 Å and Cu(4)-O(4) = 2.464Å). The Cu(2) polyhedron can also be described as a distorted octahedron (Cu(2)-4O = 1.903-2.045 Å, Cu(2)-O(8)=2.316 Å and Cu(2)-O(4) = 2.779 Å). The isotypic phosphate $Cu_4O(PO_4)_2$ (ICSD-1666) [14] (Supplementary Note 1, Table 2) has the same structure as arsenate, except that the copper ions in the phosphate are replaced as follows: Cu1 by Cu2, Cu2 by Cu3, and Cu3 by Cu1. Discrepancies in the designations of the oxygen ions in arsenate [35] and phosphate [14] were taken into account, too. Later on, we will only focus on two polymorphic modifications of the arsenate $Cu_4O(AsO_4)_2$; while similar data on the polymorphic modifications of the phosphate $Cu_4O(PO_4)_2$ isotypic to them will appear in Tables 1 and 2 (Supplementary Note 1).

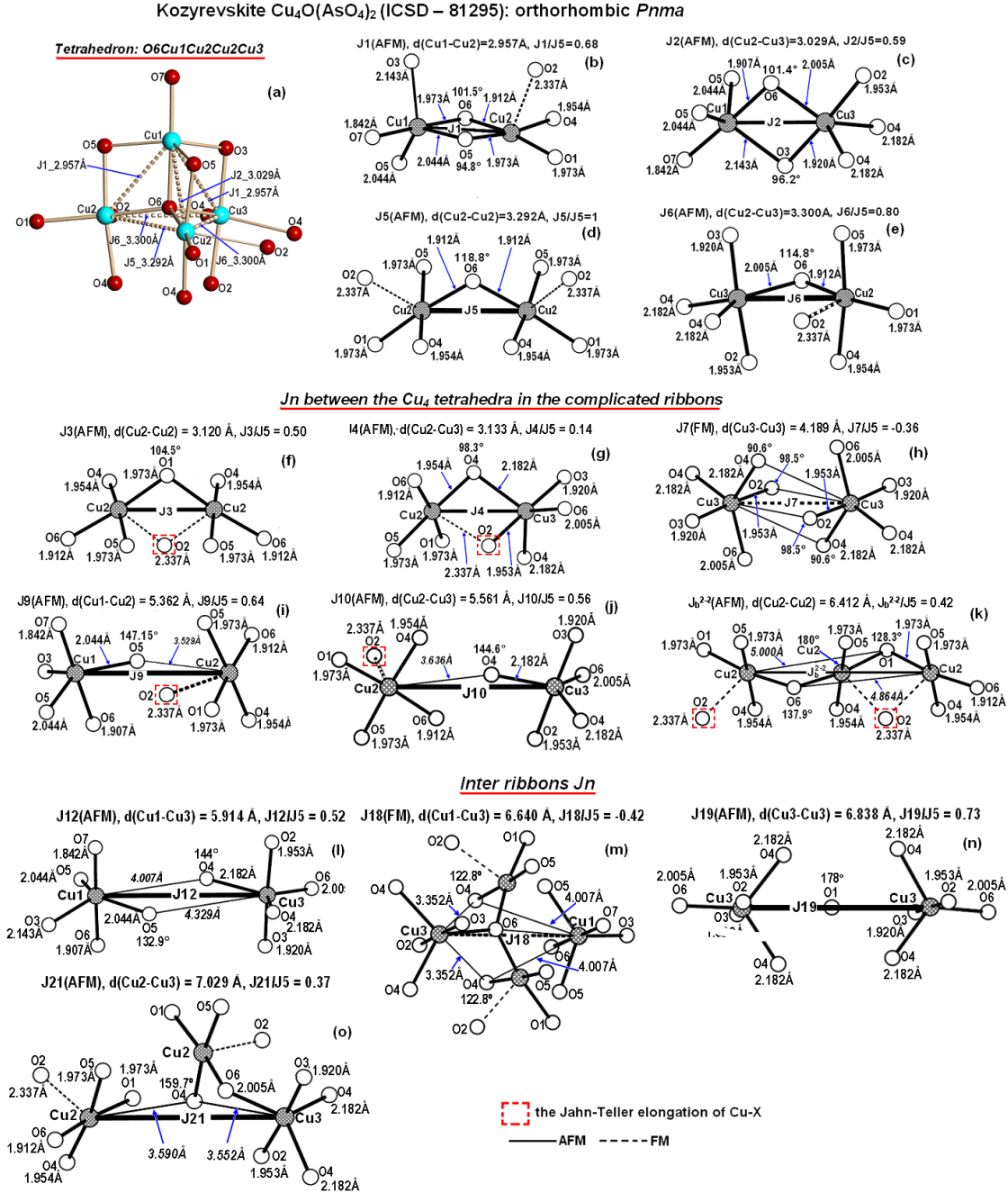

**Figure 3**. Assembly of the $CuO_m$ coordination polyhedra into tetramers I (a) in the kozyrevskite $Cu_4O(AsO_4)$. The arrangement of intermediate ions in the local space of AFM $J1$ (b), $J2$ (c), $J5$ (d) and $J6$ (e) in the tetrahedron O6Cu1Cu2Cu2Cu3 and in the local space of AFM $J3$ (f), $J4$ (g), $J9$ (i), $J10$ (j), $J_b^{2-2}$ (k) and FM $J7$ (h) between the $Cu_4$ tetrahedra in the complicated ribbons and in AFM $J12$ (l), $J19$ (n), $J21$ (o) and FM $J18$ (m) between these complicated ribbons.

After assembly of the $CuO_m$ coordination polyhedra into tetramers, the following oxocenetered $OCu_4$ tetrahedra form: O6Cu1Cu2Cu2Cu3 in the orthorhombic arsenate $Cu_4O(AsO_4)_2$ (Fig. 3a), O7Cu1Cu3Cu3Cu2 in the orthorhombic phosphate $Cu_4O(PO_4)_2$, O3Cu1Cu2Cu3Cu4 in the triclinic arsenate $Cu_4O(AsO_4)_2$ and O1Cu2Cu3Cu1Cu4 in the triclinic phosphate $Cu_4O(PO_4)_2$. Structurally, these tetrahedra do not have shared copper atoms (Fig. 4); however, as we will show below, there are strong magnetic couplings between them. In this paper, we will look at these polymorphic modifications of $Cu_4O(As(P)O_4)_2$ at a different angle and show the crystal structure of the sublattices of magnetic $Cu^{2+}$ ions.

What the crystal structures of the sublattice of the magnetic $Cu^{2+}$ ions of both modifications have in common are the complicated ribbons made up by the $Cu_4$ tetrahedra extending along the *b*-axis in *Pnma* (figures 4(a) and 5(a)) and along the *a*-axis in P -1 (figures 4(b) and 5(b)). It appears as two parallel rows of tetrahedra, with the vertices nearly meshing with each other. The main difference in the crystal structure of the complicated ribbons between the two polymorphs is that the tetrahedra of the two rows mesh into each other in a perfectly aligned manner in the rhombic modification and are misaligned in the triclinic modification. Such a displacement leads to differences in magnetic coupling (figures 4(c) and (d)).

Thus, the crystal structures of the $Cu^{2+}$ sublattice in the two distinct $Cu_4O(As(P)O_4)_2$ modifications—orthorhombic and triclinic—have quite a lot in common. However, there is no evidence about polymorphic transitions of one modification to another, which are normally caused by pressure or temperature. Transitions as these can be either reversible (enantiotropic), which, together with changes in magnetic properties, would be of particular interest, or irreversible (morphotropic). Polymorphism is common in $Cu^{2+}$-based compounds, because the JT effect makes $Cu^{2+}$ coordination be quite flexible and is responsible for there being transitional forms between the main types of coordination: (4+2), a distorted octahedron; (4+1), a tetragonal pyramid and a trigonal bipyramid; and (4), a square [25, 27].

*3.2. Structural-magnetic models*

In this chapter, we will build and discuss the structural-magnetic models of the arsenate $Cu_4O(AsO_4)_2$ in the form of its orthorhombic modification kozyrevskite and triclinic modification ericlaxmanite, and the phosphates $Cu_4O(PO_4)_2$ isotypic to them. According to our calculations (Tables 1 and 2, Supplementary Note 1), strong AFM couplings occur along all $Cu_4$ tetrahedral edges. A major contribution to the AFM components of all these couplings is made by the intermediate oxygen ions centering these tetrahedra (the O6 ion in the oxocentered tetrahedron O6Cu1Cu2Cu2Cu3 in kozyrevskite and the O3 ion in the oxocentered tetrahedron O3Cu1Cu2Cu3Cu4 in ericlaxmanite).

*3.2.1. Structural-magnetic model of the orthorhombic arsenate $Cu_4O(AsO_4)_2$ and isotypic phosphate $Cu_4O(PO_4)_2$.* Let us look closer at the characteristics that the magnetic couplings in these magnetics could possess if their formation were due to the crystal structure alone. The arrangement of intermediate ions in the local space of AFM *J*1 (figure 3(b)), *J*2 (figure 3(c)), *J*5 (figure 3(d)) and *J*6 (figure 3(e)) in tetrahedron O6Cu1Cu2Cu2Cu3 for the kozyrevskite $Cu_4O(AsO_4)_2$ is shown in figures 3 and 4(a). The strongest AFM coupling $J5$ ($J5 = -0.0789$ Å$^{-1}$, d(Cu2-Cu2) = 3.292 Å) in the tetrahedron is directed along the $b$-axis. This coupling is at the same time dominating throughout the structure. The second and third strongest couplings in the tetrahedron are two AFM couplings, $J6$ (d(Cu2-Cu3) = 3.300 Å, $J6/J5 = 0.80$), and the fourth and fifth are two AFM couplings, $J1$ (d(Cu1-Cu2) = 2.957 Å, $J1/J5 = 0.68$) and AFM $J2$ (d(Cu1-Cu3) = 3.029 Å, $J2/J5 = 0.59$). All couplings in the tetrahedron are strong antiferromagnetic couplings, which compete with each other in the triangles along the tetrahedral edges and therefore, the tetrahedra are frustrated.

In the row lying along he $b$-axis (Fig. 4(a)), these tetrahedra are coupled, at short distances, by rather strong AFM interactions $J3$ (d(Cu2-Cu2) = 3.120 Å, $J3/J5 = 0.50$) (figure 3(f)). Additionally, they are coupled, at long distances, by two strong AFM interactions $J9$ (d(Cu1-Cu2) = 5.362 Å, $J9/J5 = 0.57$) (figure 3(i)) and AFM $J_b^{2-2}$ (d(Cu2-Cu2) = 6.412 Å, $J_b^{2-2}/J5 = 0.42$) (figure 3(k)), and one weaker FM interaction, $J_b^{3-3}$ (d(Cu3-Cu3) = 6.412 Å, $J_b^{3-3}/J5 = -0.26$). All these couplings are frustrated, too, as they compose three-spin AFM systems along the linear chain $J3$-$J5$-$J_b^{2-2}$ and in the triangle $J3$-$J9$-$J1$ ($J3/J1 = 0.74$ and $J9/J5 = 0.85$). In the triangle Cu1Cu3Cu3, the ferromagnetic coupling $J_b^{3-3}$ competes with $J1$ ($J_b^{3-3}/J1 = -0.38$) and FM $J22$ (d(Cu1-Cu3) = 7.091 Å, $J22/J1 = -0.25$).

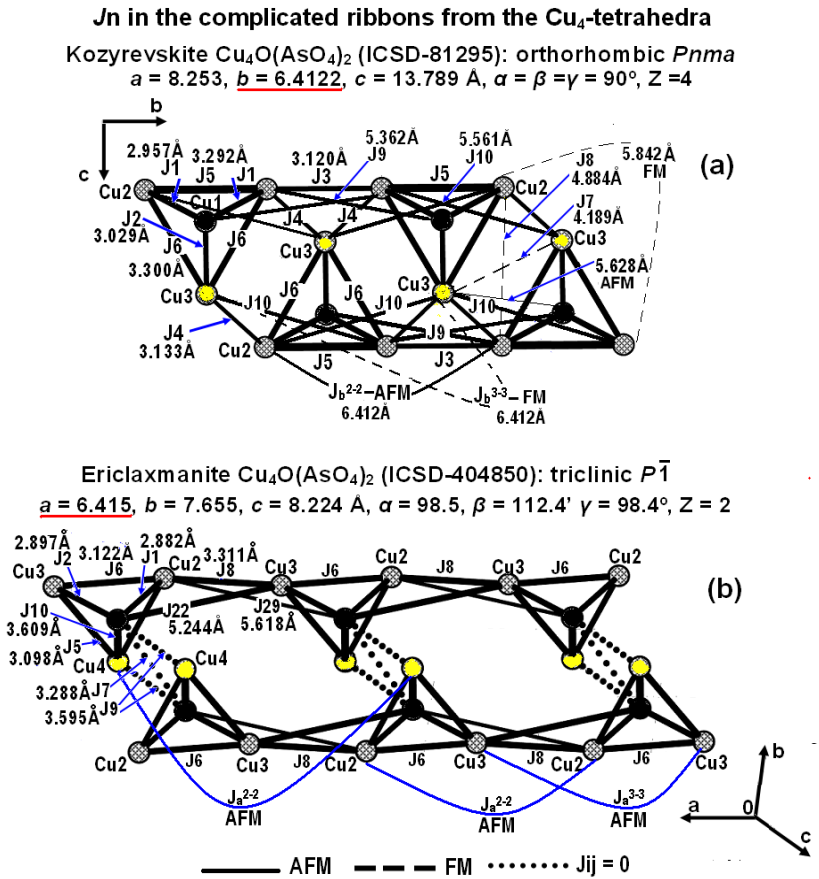

**Figure 4.** The couplings, $Jn$, in the complicated ribbons from the Cu4 tetrahedra in the kozyrevskite $Cu_4O(AsO_4)$ (*Pnma*) (a) and in the ericlaxmanite $Cu_4O(AsO_4)$ (*P* -1) (b). In this and the other figures, the thickness of lines is proportional to the strength of the couplings $Jn$. AFM and FM couplings are indicated by solid and dashed lines, respectively. Possible FM→AFM transitions are shown by strokes in the dashed lines.

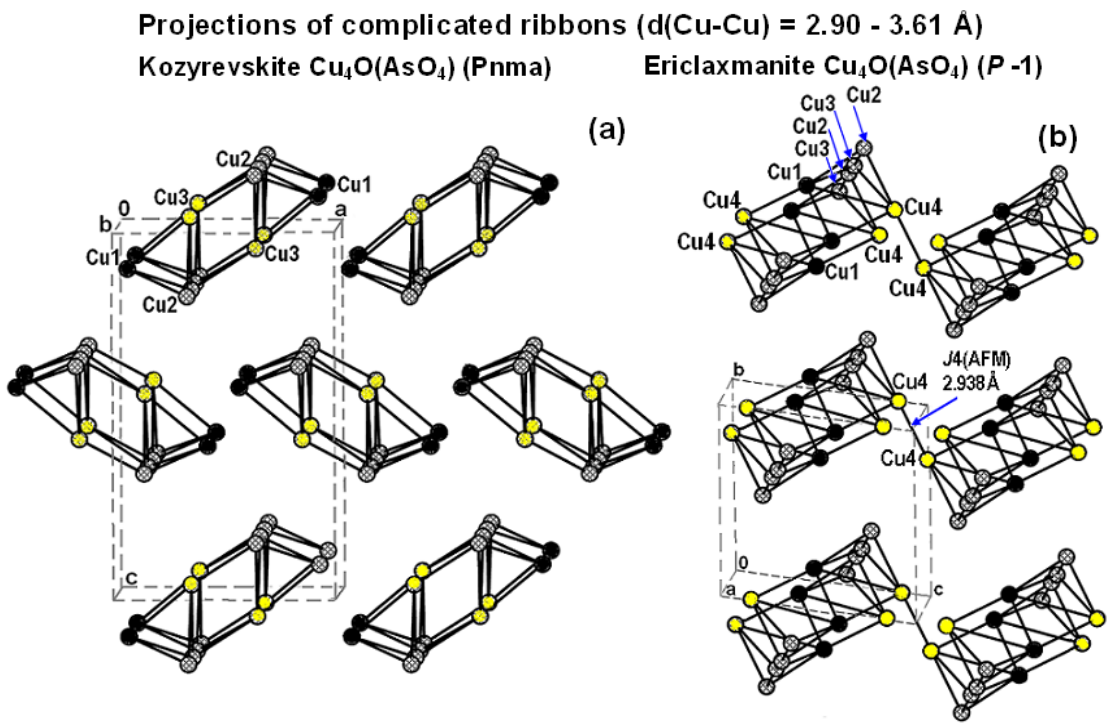

**Figure 5.** Complicated ribbons (d(Cu-Cu) = 2.90 - 3.61 Å) in projected perpendicular onto the *b*-axis in the kozyrevskite $Cu_4O(AsO_4)$ (*Pnma*) (a) and perpendicular onto the *a*-axis in the ericlaxmanite $Cu_4O(AsO_4)$ (*P* -1) (b).

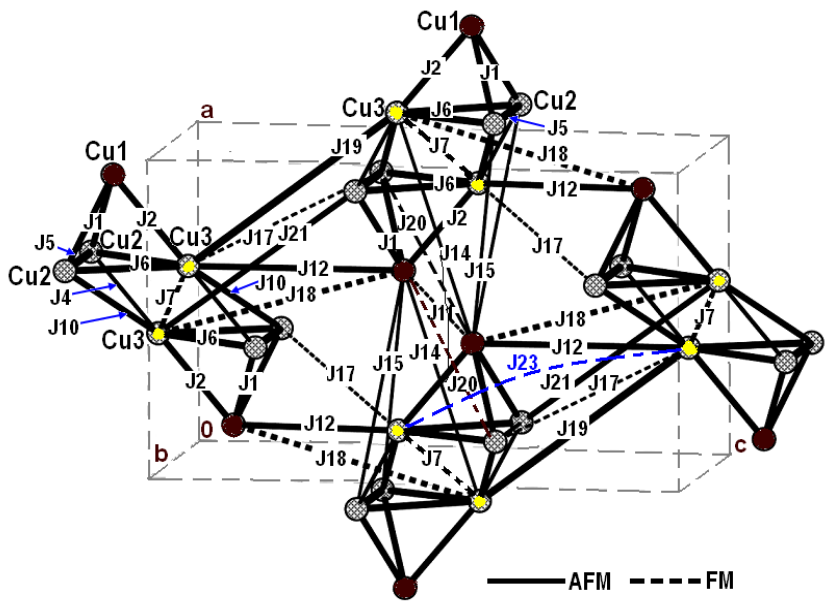

**Figure 6.** *J*n between the complicated ribbons composed of the $Cu_4$ tetrahedera in the kozyrevskite $Cu_4O(AsO_4)$ (*Pnma*).

There are two AFM couplings between the rows in the complicated ribbon: one at a short distance, but weak, $J4$ (d(Cu2-Cu3) = 3.133 Å, $J4/J5 = 0.14$) (figure 3(g)), and another at a long distance, but much stronger, $J10$ (d(Cu2-Cu3) = 5.561 Å, $J10/J5 = 0.56$) (figure 3(j)). Both these couplings are frustrated, as they compete with each other in the AFM triangles $J5$-$J4$-$J10$ and $J3$-$J4$-$J4$ ($J4/J3 = 0.28$). Three within-ribbon FM couplings—$J7$ (d(Cu3-Cu3) = 4.189 Å, $J7/J5 = -0.36$), $J8$ (d(Cu2-Cu2) = 4.884 Å, $J8/J5 = -0.06$) and $J_b^{3-3}$ (d(Cu3-Cu3) = 6.412 Å, $J_b^{3-3}/J5 = -0.26$) — are not competing with within-ribbon interactions. However, they are active players in the competition with between-ribbon interactions. The strongest interactions between the complicated ribbons (figure 6) are $J19$ (d(Cu3-Cu3) = 6.838 Å, $J19/J5 = 0.73$) (figure 3(n)), AFM $J12$ (d(Cu1-Cu3) = 5.914 Å, $J12/J5 = 0.52$) (figure 3(l)), FM $J18$ (d(Cu1-Cu3) = 6.640 Å, $J18/J5 = -0.42$) (figure 3(m)) and $J21$ (d(Cu2-Cu3) = 7.029 Å, $J21/J5 = 0.38$) (figure 3(o)). The couplings between the complicated ribbons (figure (6)) form six triangles of two types: AFM-AFM-AFM (figure 1(a)) and AFM-FM-FM (figure 1(b)), and thus introduce an additional competition to the complicated ribbons. The competition between the AFM couplings $J5$-$J15$-$J15$ ($J15/J5 = 0.27$) in the triangles Cu1Cu2Cu2 and between the AFM couplings $J6$-$J14$-$J15$ in the triangles Cu1Cu2Cu3 is type 1 competition ($J14/J6 = 0.31$, $J15/J6 = 0.33$). The competition in the following triangles is type 2 competition: (1) in Cu3Cu3Cu3, between AFM$J19$–FM$J7$–FM $J22$ ($J7/J19 = -0.54$, $J22/J19 = 0.32$), (2) in Cu1Cu2Cu1, between AFM$J1$–FM$J11$–FM$J20$ ($J11/J1 = -0.32$, $J20/J1 = -0.32$) and (3) in Cu1Cu3Cu3, between AFM$J12$–FM$J18$–FM7 ($J18/J12 = -0.80$, $J7/J12 = -0.69$) and AFM$J21$–FM$J7$–FM$J17$ ($J7/J21 = -0.95$, $J17/J21 = -0.63$)).

Thus, according to our calculations, the magnetic structures of the orthorhombic arsenate $Cu_4O(AsO_4)_2$ and isotypic phosphate $Cu_4O(PO_4)_2$ are strongly frustrated antiferromagnetics. Their base elements are the complicated ribbons composed of AFM $Cu_4$ tetrahedra directed along the *b*-axis and coupled by rather strong AFM and FM interactions. All magnetic interactions within and between the complicated ribbons are frustrated.

*3.2.2. Structural-magnetic model of the triclinic modification of the arsenate $Cu_4O(AsO_4)_2$ and isotypic phosphate $Cu_4O(PO_4)_2$.* An increase in crystallographically independent Cu sites up to up to four and a many-fold increase in spin-spin interactions, $J_{ij}$, in a low-symmetry magnetic system in ericlaxmanite leads to a large number of frustrations. To identify all frustrated fragments appearing as triangles and linear chains in this complex system, we calculated 70 existing spin-spin interactions in the triclinic arsenate $Cu_4O(AsO_4)_2$ and isotypic phosphate $Cu_4O(PO_4)_2$2 not only at short, but also at long distances (figures 4(b), and 7); Supplementary Note 1, Table 2 and figures 4(b), 7(a) and 7(b)).

As we demonstrated above, what the two polymorphs—the orthorhombic and the triclinic modification—have in common is two rows of separate tetrahedra, which form complicated ribbons. In the triclinic modification (ericlaxmanite), they are directed along the *a*-axis (figures 4(b) and 5(b)). In ericlaxmanite, the oxocentered tetrahedron O3Cu1Cu2Cu3Cu4 is strongly distorted. Along the six edges, the strength of AFM coupling increases (from -0.0335 Å$^{-1}$ to -0.1154 Å$^{-1}$) with an increase in edge length (from 2.882 Å to 3.609 Å) and the corresponding increase of the angle CuO3Cu (from 97.6° to 140.5°) (Supplementary Note 1, Table 2).

The dominating coupling in ericlaxmanite is the AFM coupling $J10$ ($J10$ = -0.1154 Å$^{-1}$, d(Cu1-Cu4) = 3.609 Å) along the longest edge Cu1-Cu4 of the tetrahedron Cu1Cu2Cu3Cu4, with the O3 ion ($j$(O3) = -0.1154 Å$^{-1}$) centering this tetrahedron being the main contributor to the AFM component. This coupling ($J10/J5$=1.46) is stronger than the dominating coupling $J5$ in kozyrevskite. The strengths of AFM coupling in the tetrahedron are as follows (in descending order): $J6$ (d(Cu2-Cu3) = 3.122 Å, $J6/J10$ = 0.54), $J5$ (d(Cu3-Cu4) = 3.098 Å, $J5/J10$ = 0.51), $J3$ (d(Cu2-Cu4) = 2.927 Å, $J3/J10$ = 0.41), $J2$ (d(Cu1-Cu3) = 2.897 Å, $J2/J10$ = 0.41) and $J1$ (d(Cu1-Cu2) = 2.882 Å, $J1/J10$ = 0.29) (figure 4(b)).

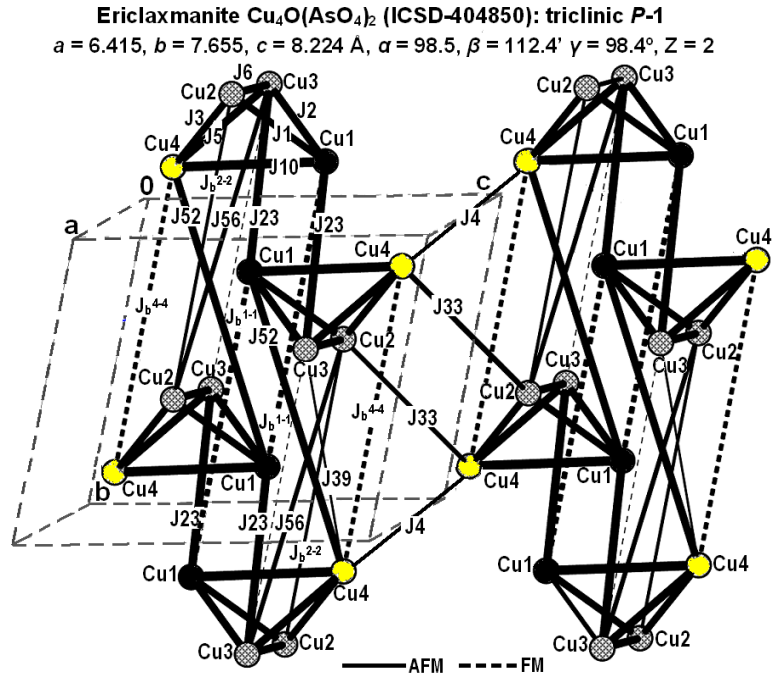

**Figure 7.** $J$n between the complicated ribbons from the Cu4-tetrahedera in the triclinic ($P$ -1) ericlaxmanite Cu$_4$O(AsO$_4$).

In the row, these tetrahedra are coupled, at short distances, by quite strong AFM interactions $J8$ (d(Cu2-Cu3) = 3.311 Å, $J8/J10$ = 0.56), similar to $J4$ in kozyrevskite. Furthermore, at longer distances, they are additionally coupled by strong AFM interaction $J22$ (d(Cu1-Cu3) = 5.244 Å, $J22/J10$ = 0.54) and weak AFM $J29$ (d(Cu1-Cu2) = 5.618 Å, $J29/J10$ = 0.10). A 20-fold increase in the strength of AFM coupling $J21$ ($J21$ = -0.0423 ↔ -0.0021 Å$^{-1}$, d(Cu3-Cu4) = 5.211 Å, $J21/J10$ = 0.37 ↔ 0.02) between Cu3 and Cu4 ions is possible because of a minor displacement of the intermediate ion O5. Immediately along the parameter $a$, along which the rows of tetrahedra lie in the triclinic modification, the strength of the AFM coupling $J_a^{1-1}$ grew 2.4-fold ($J_a^{1-1}$ = -0.0205 Å$^{-1}$, d(Cu1-Cu1) = 6.415 Å, $J_a^{1-1}/J10$ = 0.18) compared to this variable in the orthorhombic modification $J_b^{1-1}$ ($J_b^{1-1}$ = 0.0086 Å$^{-1}$, d(Cu1-Cu1) = 6.412 Å, $J_b^{1-1}/J5$ = -0.11).

By contrast, the strong AFM coupling $J_b^{2-2}$ ($J_b^{2-2}$ = -0.0329 Å$^{-1}$, d(Cu2-Cu2) = 6.412 Å, $J_b^{2-2}/J5$ = 0.42) between the Cu2-Cu2 ions that exist in the orthorhombic modification split into two AFM couplings with reduced strengths in the triclinic modification: $J_a^{2-2}$ ($J_a^{2-2}$ = -0.0210 Å$^{-1}$, d(Cu2-Cu2) = 6.415 Å, $J_a^{2-2}/J10$ = 0.18) and $J_a^{3-3}$ ($J_a^{3-3}$ = -0.0215 Å$^{-1}$, d(Cu2-Cu2) = 6.415 Å, $J_a^{3-3}/J10$ = 0.19).

The most substantial changes in magnetic coupling at lower symmetry took places in the complicated ribbon due to a misalignment of the rows (figures 4(b) and (d)). First, closely spaced tetrahedra from different rows formed pairs. Magnetic coupling between the tetrahedra in these pairs vanished ($J7 = 0$, d(Cu1-Cu1) = 3.288 Å; $J9 = 0$, d(Cu1-Cu4) = 3.595 Å) or became ferromagnetic (FM $J40 = 0.0224$ Å$^{-1}$, d(Cu4-Cu4) = 6.410 Å). Two strong AFM couplings, $J23$ ($J23 = -0.0707$ Å$^{-1}$, Cu1-Cu3 = 5.299 Å) and AFM $J42$ ($J4 = -0.0224$ Å$^{-1}$, d(Cu3-Cu4) = 6.482 Å), which compete with AFM $J10$ in the triangle Cu1Cu4Cu3, appeared instead (figures 4(b) and 7(a)). Secondly, in addition to these couplings, strong FM couplings, $J57$, appeared within the ribbon ($J57 = 0.0285$ Å$^{-1}$, Cu4-Cu4 = 7.787 Å) (figure 4(b)), which compete in the triangle Cu1Cu4Cu4 with the strong between-ribbon FM couplings $J53$ ($J53 = 0.0185$ Å$^{-1}$, d(Cu1-Cu4) = 7.517 Å) and AFM $J52$ ($J52 = -0.0338$ Å$^{-1}$, d(Cu1-Cu4) = 7.467 Å). In the triangle Cu3Cu4Cu4, they compete with the weaker between-ribbon FM couplings $J47$ ($J47 = 0.0068$ Å$^{-1}$, d(Cu3-Cu4) = 7.197 Å) and AFM $J39$ ($J39 = -0.0083$ Å$^{-1}$, d(Cu3-Cu4) = 6.370 Å).

The interactions between the complicated ribbons (Supplementary: figures 4(b), 7(a) and 7(b)) are weaker than the strongest interactions within the ribbon. The AFM coupling $J48$ (d(Cu1-Cu3) = 7.213 Å, $J48/J10 = 0.42$) is the strongest. The strengths of the top 20 strongest AFM interactions between the complicated ribbons range from -0.0128 Å$^{-1}$ to -0.0484 Å$^{-1}$. There are much fewer FM couplings between the complicated ribbons and they are weaker than the AFM couplings. The strengths of the top six strongest FM couplings range from 0.0146 Å$^{-1}$ to 0.0227 Å$^{-1}$. Additionally, even a minor displacement of the intermediate copper ion in the local space between the magnetic ions Cu2-Cu4 affects the strength of two FM couplings, $J46$ ($J46 = 0.0089$ Å$^{-1}$ ↔ 0.0299 Å$^{-1}$, d(Cu2-Cu4) = 7.050 Å) and $J54$ ($J54 = 0.0068$ Å$^{-1}$ ↔ 0.0263 Å$^{-1}$, d(Cu2-Cu4) = 7.596 Å). In $J37$ ($J37 = 0.0089$ Å$^{-1}$ FM ↔ -0.0159 Å$^{-1}$ AFM, d(Cu2-Cu4) = 6.243 Å), a displacement of the intermediate ion O1 in the local space between the magnetic ions Cu1-Cu2 can lead to an FM ↔ AFM transition.

All magnetic couplings, both in the complicated ribbons and between them, are frustrated. The parameters of the frustrated triangles are given in Table 3.

*3.2.3. Frustration of antiferromagnetic $Cu_4O(As(P)O_4)_2$.* As far as the minerals in question are considered, in which separate magnetic fragments in the form of AFM Cu$_4$ tetrahedra assembled into complicated ribbons, with AFM coupling between them, it is not easy to understand the hierarchy of magnetic interactions. In addition to the nearest interactions in the frustrated AFM tetrahedral ribbons, even very weak long-range interactions, including those between the ribbons, can too have a strong impact on the magnetic state of a frustrated quantum magnet at low temperatures [37, 38].

Based on our calculations, we conclude that three main structural factors define the magnetic spin-frustrated systems of the anti-ferromagnetics $Cu_4O(As(P)O_4)_2$ with strongly entangled competing magnetic interactions in the sublattice of magnetic ions $Cu^{2+}$ (Table 3, figures 4 and 7). First, the existence of frustrated oxocentered AFM OCu$_4$ tetrahedra ("soft islands"). Secondly, competition between a large number of long-range secondary couplings, both strong enough and weak with frustrated primary couplings due to Cu$_4$ tetrahedra. Finally, an additional competition between long-range secondary communications themselves. The competition continues in both triangular and linear systems with three spins, AFM-AFM-AFM and AFM-FM-FM.

As both spatial groups—orthorhombic *Pnma* (N62) and triclinic *P* –1 (N2), in which $Cu_4O(As(P)O_4)_2$ crystallize—are centrosymmetric, it might be expected that all magnetic ions in them are related by the inversion center; however, they do not live up to this expectation. In the orthorhombic modification of $Cu_4O(AsO_4)_2$, the copper ions occupy three crystallographically independent sites (Cu1, Cu2, and Cu3) and only the Cu2 ions are in the centrosymmetric equivalent position 8d, while the Cu1 and Cu3 ions are in the non-centrosymmetric equivalent position 4c, in which case the antisymmetric anisotropic exchange interaction (the Dzyaloshinskii-Moriya interaction) [39, 40] contributes to the total magnetic exchange interaction between two nearest magnetic spins in the lattice ($D \neq 0$ in the Dzyaloshinskii-Moriya interaction). The competition between the exchange interaction and the Dzyaloshinskii-Moriya interaction caused by spin-orbital coupling induces spin canting and, thus, serves as a source of a weak FM behavior in an AFM. The results of a magnetic susceptibility measurement made on orthorhombic $Cu_4O(AsO_4)_2$ by Adams et al. [33] are consistent with these structural data.

**Table 3.** Parameters of the frustrated triangles in the triclinic modification of the arsenate $Cu_4O(AsO_4)_2$

| Frustrated triangles | $Jn$ | Bond | d(Cu-Cu) (Å) | $Jn^{(a)}$ (Å$^{-1}$) | $Jn/Jn^{max}$ |
|---|---|---|---|---|---|
| *In tetrahedron* | | | | | |
| Cu1Cu2Cu3 | $J6$ | Cu2 - Cu3 | 3.122 | -0.0624 (AFM) | 1 |
| | $J2$ | Cu1 - Cu3 | 2.897 | -0.0469 (AFM) | 0.75 |
| | $J1$ | Cu1 - Cu2 | 2.882 | -0.0335 (AFM) | 0.54 |
| Cu1Cu2Cu4 | $J10$ | Cu1 -Cu4 | 3.609 | -0.1154 (AFM) | 1 |
| | $J3$ | Cu2 - Cu4 | 2.927 | -0.0474 (AFM) | 0.41 |
| | $J1$ | Cu1 - Cu2 | 2.882 | -0.0335 (AFM) | 0.29 |
| Cu1Cu3Cu4 | $J10$ | Cu1 - Cu4 | 3.609 | -0.1154 (AFM) | 1 |
| | $J5$ | Cu3 - Cu4 | 3.098 | -0.0588 (AFM) | 0.51 |
| | $J2$ | Cu1 - Cu3 | 2.897 | -0.0469 (AFM) | 0.41 |
| Cu2Cu3Cu4 | $J6$ | Cu2 - Cu3 | 3.122 | -0.0624 (AFM) | 1 |
| | $J5$ | Cu3 - Cu4 | 3.098 | -0.0588 (AFM) | 0.94 |
| | $J3$ | Cu2 - Cu4 | 2.927 | -0.0474 (AFM) | 0.76 |
| *In complicated ribbons between tetrahedra* | | | | | |
| Cu1Cu4Cu3 | $J10$ | Cu1-Cu4 | 3.609 | -0.1154 (AFM) | 1 |
| | $J23$ | Cu1-Cu3 | 5.299 | -0.0707 (AFM) | 0.62 |
| | $J42$ | Cu3-Cu4 | 6.482 | -0.0224 (AFM) | 0.19 |
| Cu4Cu1Cu3 | $J10$ | Cu1-Cu4 | 3.609 | -0.1154 (AFM) | 1 |
| | $J22$ | Cu1-Cu3 | 5.244 | -0.0625 (AFM) | 0.54 |
| | $J21$ | Cu4-Cu3 | 5.211 | -0.0423Å (AFM) ↔-0.0021(AFM) | 0.37 |
| Cu1Cu2Cu3 | $J8$ | Cu2-Cu3 | 3.311 | -0.0642 (AFM) | 1 |
| | $J22$ | Cu1-Cu3 | 5.244 | -0.0625 (AFM) | 0.97 |
| | $J1$ | Cu1-Cu2 | 2.882 | -0.0335 (AFM) | 0.52 |
| Cu1Cu2Cu1 | $J1$ | Cu1-Cu2 | 2.882 | -0.0335 (AFM) | 1 |
| | $Ja^{1-1}$ | Cu1-Cu1 | 6.415 | -0.0205 (AFM) | 0.61 |
| | $J29$ | Cu2-Cu1 | 5.618 | -0.0118 (AFM) | 0.35 |
| Cu2Cu3Cu1 | $J8$ | Cu2-Cu3 | 3.311 | -0.0642 (AFM) | `1 |
| | $J2$ | Cu3-Cu1 | 2.897 | -0.0469 (AFM) | 0.73 |
| | $Ja^{1-1}$ | Cu1-Cu1 | 6.415 | -0.0205 (AFM) | 0.32 |
| Cu2Cu3Cu1 | $J8$ | Cu2-Cu3 | 3.311 | -0.0642 (AFM) | 1 |
| | $J2$ | Cu3-Cu1 | 2.897 | -0.0469 (AFM) | 0.73 |
| | $J29$ | Cu2-Cu1 | 5.618 | -0.0118 (AFM) | 0.18 |
| *Between complicated ribbons* | | | | | |
| Cu2Cu2Cu3 | $J6$ | Cu2 - Cu3 | 3.122 | -0.0624 (AFM) | 1 |
| | $J56$ | Cu2 - Cu3 | 7.618 | -0.0178 (AFM) | 0.29 |
| | $J_b^{2-2}$ | Cu4 - Cu4 | 7.655 | -0.0095 (AFM) | 0.15 |
| Cu1Cu2Cu3 | $J6$ | Cu2 - Cu3 | 3.122 | -0.0624 (AFM) | 1 |
| | $J16$ | Cu1 - Cu3 | 4.522 | -0.0224 (AFM) | 0.36 |
| | $J38$ | Cu1 – Cu2 | 6.245 | -0.0171(AFM) | 0.27 |
| Cu4Cu4Cu3 | $J5$ | Cu3 - Cu4 | 3.098 | -0.0588 (AFM) | 1 |
|  | $J_c^{4-4}$ | Cu4 - Cu4 | 8.224 | -0.0176 (AFM) | 0.30 |
|  | $J36$ | Cu3 - Cu4 | 6.210 | -0.0141 (AFM) | 0.34 |
| Cu1Cu3Cu1 | $J48$ | Cu1 - Cu3 | 7.213 | -0.0484 (AFM) | 1 |
|  | $J16$ | Cu1 - Cu3 | 4.522 | -0.0224 (AFM) | 0.46 |
|  | $J45$ | Cu1 - Cu1 | 6.854 | -0.0128 (AFM) | 0.26 |
| Cu2Cu4Cu4 | $J3$ | Cu2 - Cu4 | 2.927 | -0.0474 (AFM) | 1 |
|  | $J33$ | Cu2 - Cu4 | 6.143 | -0.0178 (AFM) | 0.38 |
|  | $J58$ | Cu4 - Cu4 | 8.147 | -0.0108 (AFM) | 0.23 |
| Cu2Cu4Cu4 | $J3$ | Cu2 - Cu4 | 2.927 | -0.0474 (AFM) | 1 |
|  | $J4$ | Cu4 - Cu4 | 2.938 | -0.0280 (AFM) | 0.59 |
|  | $J30$ | Cu2 - Cu4 | 5.853 | -0.0254 (AFM) | 0,54 |
| Cu1Cu3Cu4 | $J2$ | Cu1 - Cu3 | 2.897 | -0.0469 (AFM) | 1 |
|  | $J52$ | Cu1 - Cu4 | 7.467 | -0.0338 (AFM) | 0.72 |
|  | $J39$ | Cu3 - Cu4 | 6.370 | -0.0083 (AFM) | 0.18 |
| Cu1Cu1Cu2 | $J1$ | Cu1 - Cu2 | 2.882 | -0.0335 (AFM) | 1 |
|  | $J_c^{1-1}$ | Cu1 - Cu1 | 8.224 | -0.0178 (AFM) | 0.53 |
|  | $J37$ | Cu1 - Cu2 | 6.243 | -0.0159 (AFM) ↔0.0089 FM | 0.47 |
| Cu1Cu3Cu4 | $J16$ | Cu1 - Cu3 | 4.522 | -0.0224 (AFM) | 1 |
|  | $J39$ | Cu3 - Cu4 | 6.370 | -0.0083 (AFM) | 0.37 |
|  | $J27$ | Cu1 - Cu4 | 5.561 | -0.0074 (AFM) | 0.33 |
| Cu1Cu2Cu2 | $J1$ | Cu1 - Cu2 | 2.882 | -0.0335 (AFM) | 1 |
|  | $J_c^{2-2}$ | Cu2 - Cu2 | 8.224 | -0.0215 (AFM) | 0.64 |
|  | $J37$ | Cu1 - Cu2 | 6.243 | -0.0159 (AFM) ↔0.0089 (FM) | 0.47 |
| Cu1Cu4Cu4 | $J52$ | Cu1 – Cu4 | 7.467 | -0.0338 (AFM) | 1 |
|  | $J57$ | Cu4 - Cu4 | 7.787 | 0.0285 (FM) | -0.84 |
|  | $J53$ | Cu1 - Cu4 | 7.517 | 0.0185 (FM) | -0.55 |
| Cu1Cu2Cu3 | $J1$ | Cu1 – Cu2 | 2.882 | -0.0335 (AFM) | 1 |
|  | $J49$ | Cu2 – Cu3 | 7.316 | 0.0167 (FM) | -0.50 |
|  | $J18$ | Cu1 – Cu3 | 4.742 | 0.0071 (FM) | -0.21 |
| Cu3Cu4Cu4 | $J57$ | Cu4 - Cu4 | 7.787 | 0.0285 (FM) | 1 |
|  | $J39$ | Cu3 - Cu4 | 6.370 | -0.0083 (AFM) | -0.29 |
|  | $J47$ | Cu3 – Cu4 | 7.197 | 0.0068 (FM) | 0.24 |
| Cu4Cu2Cu2 | $J54$ | Cu2-Cu4 | 7.596 | 0.0263 (FM) ↔ 0.0068 (FM) | 1 |
|  | $J33$ | Cu2 - Cu4 | 6.143 | -0.0178 (AFM) | -0.68 |
|  | $J14$ | Cu2-Cu2 | 4.439 | 0.0084 (FM) | 0.032 |
| Cu1Cu1Cu3 | $J16$ | Cu1 - Cu3 | 4.522 | -0.0224 (AFM) | 1 |
|  | $J_b^{1-1}$ | Cu1 - Cu1 | 7,655 | 0.0209 (FM) | -0.93 |
|  | $J18$ | Cu1 - Cu3 | 4.742 | 0.0071 (FM) | -0.32 |
| Cu4Cu4Cu4 | $J40$ | Cu4 - Cu4 | 6.410 | 0.0224 (FM) | 1 |
|  | $J_c^{4-4}$ | Cu4 - Cu4 | 8.224 | -0.0176 (AFM) | -0.79 |
|  | $J25$ | Cu4 - Cu4 | 5.411 | 0.0044 (FM) | 0.20 |
| Cu2Cu2Cu2 | $J_c^{2-2}$ | Cu2 - Cu2 | 8.224 | -0.0215 (AFM) | 1 |
|  | $J14$ | Cu2 - Cu2 | 4.439 | 0.0084 (FM) | -0.39 |
|  | $J20$ | Cu2 - Cu2 | 5.122 | 0.0078 (FM) | -0.36 |

Significant antiferromagnetic coupling was revealed by variable temperature measurements. At very low temperatures (15-25 K), the sample undergoes a transition to a weak ferromagnetic state. At still lower temperatures, this will lead to magnetic ordering.

Importantly, the low-symmetry triclinic spatial group ($P$ -1) has only one equivalent position, 2i, and it has the inversion center. Each of the four crystallographically nonequivalent copper ions will be forced to take this position, preventing DM ordering and keeping the magnetic interactions in the triclinic modification of $Cu_4O(As(P)O_4)_2$ frustrated at lower temperatures.

## 4. Conclusions

We have built and analyzed the structural-magnetic models of two minerals from the Tolbachik volcano: (1) the arsenate $Cu_4O(AsO_4)_2$ in its orthorhombic modification (kozyrevskite) and triclinic modification (ericlaxmanite) and (2) the isotypic phosphates $Cu_4O(PO_4)$; and revealed the main correlations between the structure and magnetic properties within them. Although the chemical formula—$Cu_4O(AsO_4)_2$—is easy, the magnetic couplings in these antiferromagnetics have been found to be intricately entangled, especially in the triclinic modification, due to an increased number of crystallographically independent Cu sites (up to four) and a many-fold increase in the number of the spin-spin interactions $J_{ij}$ in a low-symmetry magnetic system.

The basic elements of both modifications are complicated ribbons composed of AFM $Cu_4$ tetrahedra. These tetrahedra have no shared copper atoms; however, there are strong anti-ferromagnetic and ferromagnetic couplings, both within the complicated ribbons and between them. It has been established that both modifications are strongly frustrated 3-D magnets due to competition between the nearest AFM interactions along the AFM $Cu_4$ tetrahedral edges and competition between these interactions and a multiplicity of long-range secondary AFM and FM interactions. Additionally, there are multiple competitions among weaker long-range interactions.

However, the state of the frustrated spin system of the orthorhombic modification (*Pnma*) is not the same as the state of the triclinic modification ($P$ -1). The main difference between them is that the centrosymmetric orthorhombic system can induce the ordering DM interaction in some interactions between copper ions in non-centrosymmetric partial positions and suppress frustration at lower temperature, which the triclinic system cannot. The DM interaction in the centrosymmetric triclinic modification of $Cu_4O(As(P)O_4)_2$ is impossible, because all copper ions are in the centrosymmetric position. It is possible that the triclinic modification of $Cu_4O(As(P)O_4)_2$ is a spin liquid.

**Data availability statement**
All data that support the findings of this study are included within the article (and any supplementary files).

**Additional information**
Supplementary information is available in the online version of the paper.

**Acknowledgments**
The work was financially supported within the frames of the State Order of the Institute of Chemistry FEBRAS, project No. 0205-2022-0001.

**ORCID iDs**
L M Volkova https://orcid.org/0000-0002-6316-8586

*Supplementary Material*

**Orthorhombic and triclinic modifications of the arsenate Cu$_4$O(AsO$_4$)$_2$ and isotypic phosphates Cu$_4$O(PO$_4$)$_2$: strongly frustrated antiferromagnetics**

*L M Volkova*

**Supplementary Note 1:**
**Figure 4 (all magnetic interactions (Jn) are shown)**

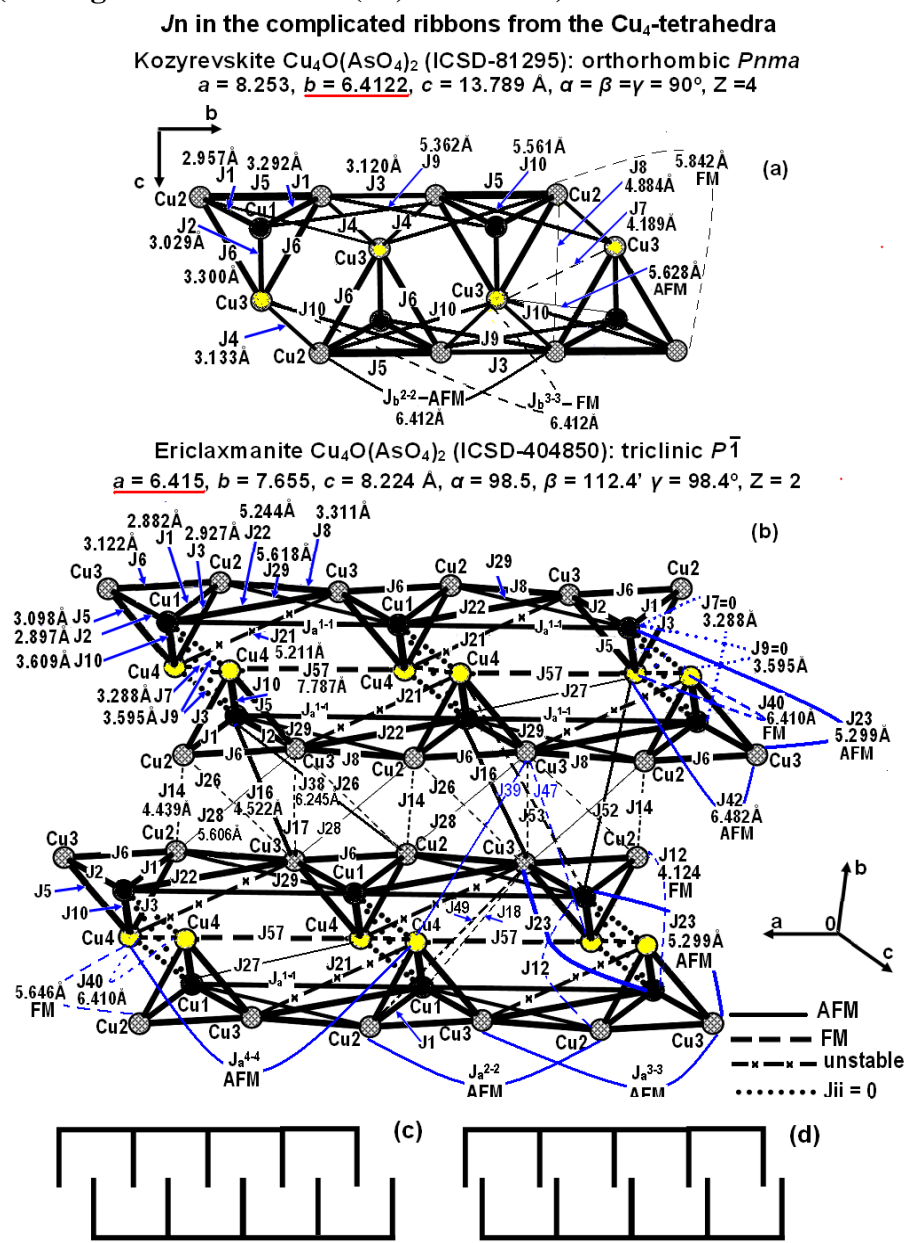

**Figure 4.** The couplings, $J$n, in the complicated ribbons from the Cu4 tetrahedra in the kozyrevskite Cu$_4$O(AsO$_4$) (*Pnma*) (a) and in the ericlaxmanite Cu$_4$O(AsO$_4$) (*P* -1) (b). In this and the other figures, the thickness of lines is proportional to the strength of the couplings $J$n. AFM and FM couplings are indicated by solid and dashed lines, respectively. Possible FM→AFM transitions are shown by strokes in the dashed lines. The diagrams complicated ribbons in the orthorhombic (c) and in the triclinic (d) modifications.

**Figure 7** (all magnetic interactions (Jn) are shown.)

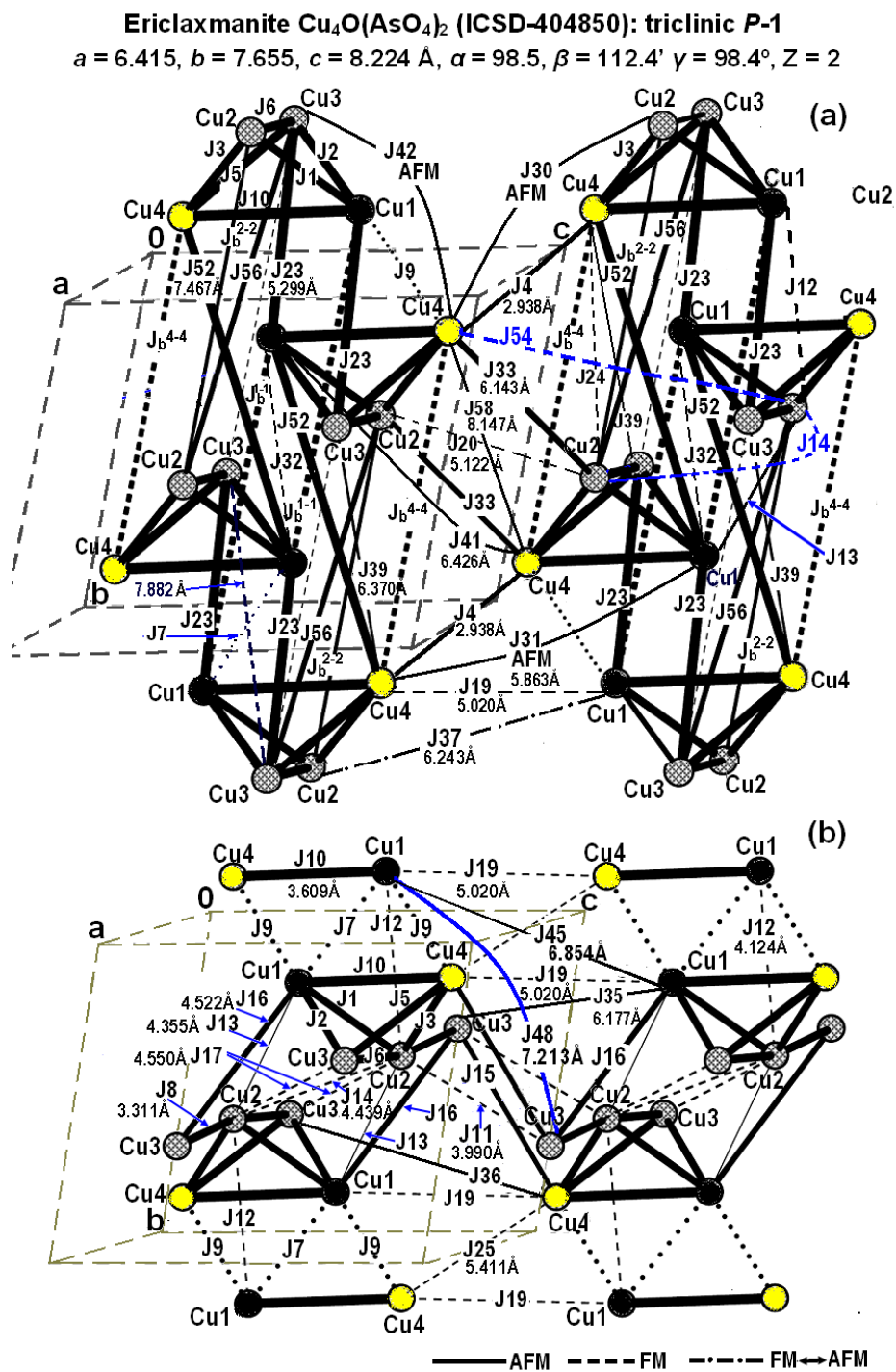

**Figure 7.** $J$n between the complicated ribbons from the Cu4-tetrahedera in the triclinic ($P$-1) ericlaxmanite $Cu_4O(AsO_4)$ (a) and (b).

**Table 1.** Crystallographic characteristics and parameters of magnetic couplings ($J$n) calculated on the basis of structural data and respective distances between magnetic $Cu^{2+}$ ions in the copper(II) oxide arsenate kozyrevskite $Cu_4O(AsO_4)_2$ and phosphate $Cu_4O(PO_4)_2$

| Crystallographic and magnetic parameters | $Cu_4O(AsO_4)$[33] Min Name: Kozyrevskite (Data for ICSD-81295) Space group *Pnma* (N62) $a = 8.253, b = 6.4122, c = 13.789$ Å $\alpha = \beta = \gamma = 90°$, Z = 4 Method[a]: XDS (293 K); $R$-value[b] = 0.038 | $Cu_4O(PO_4)$[15] (Data for ICSD-50459) Space group *Pnma* (N62) $a = 8.088, b = 6.270, c = 13.3839$ Å $\alpha = \beta = \gamma = 90°$, Z = 4 Method[a]: XDS (293 K); $R$-value[b] = 0.0338 |
|---|---|---|
| d(Cu-X) (Å) | Cu1: trigonal bipyramid<br>Cu1-O7 = 1.842 (ax)<br>-O6 = 1.907 (ax)<br>-O5 = 2.044 x 2 (eq)<br>-O3 = 2.143 (eq)<br>Cu2: tetragonal pyramid<br>Cu2-O6 = 1.912<br>-O4 = 1.954<br>-O1 = 1.973<br>-O5 = 1.973<br>-O2 = 2.337 (ax)<br>Cu3: trigonal bipyramids<br>Cu3-O3 = 1.920<br>-O2 = 1.953<br>-O6 = 2.005<br>-O4 = 2.182 x 2 | Cu1: trigonal bipyramid<br>Cu1-O1 = 1.835 (ax)<br>-O7 = 1.873 (ax)<br>-O6 = 2.105 x 2 (eq)<br>-O4 = 2.161 (eq)<br>Cu3: tetragonal pyramid<br>Cu3-O7 = 1.890<br>-O5 = 1.964<br>-O3 = 1.973<br>-O6 = 1.996<br>-O2 = 2.340 (ax)<br>Cu2: trigonal bipyramids<br>Cu2-O4 = 1.921<br>-O2 = 1.946<br>-O7 = 1.989<br>-O3 = 2.174 x 2 |
| ***Complicated ribbon*** | | |
| Bond | Cu1-Cu2 | Cu1-Cu3 |
| d(Cu-Cu) (Å) | 2.957 | 2.965 |
| $J1$[c] (Å$^{-1}$) | $J1$ = -0.0533 (AFM) | $J1$ = -0.0514 (AFM) |
| $j(X)$[d] (Å$^{-1}$) | $j$(O6): -0.0439 | $j$(O7): -0.0550 |
| ($\Delta h(X)$[e] Å, $l_n'/l_n$[f], CuXCu[g]) | (-0.192, 1.0, 101.5°) | (-0.242, 1.1, 104.0°) |
| $j(X)$[d] (Å$^{-1}$) | $j$(O5): -0.0094 | $j$(O6): 0.0036 |
| ($\Delta h(X)$[e] Å, $l_n'/l_n$[f], CuXCu[g]) | (-0.041, 1.07, 94.8°) | (0.016, 1.1, 92,6°) |
| $J$n/$J$max | $J1/J5$ = 0.68 | $J1/J5$ = 0.66 |
| Bond | Cu1-Cu3 | Cu1-Cu2 |
| d(Cu-Cu) (Å) | 3,029 | 3.012 |
| $J2$[c] (Å$^{-1}$) | $J2$ = -0.0465 (AFM) | $J2$ = -0.0488 (AFM) |
| $j(X)$[d] (Å$^{-1}$) | $j$(O6): -0.0356 | $j$(O7): -0.0426 |
| ($\Delta h(X)$[e] Å, $l_n'/l_n$[f], CuXCu[g]) | (-0.163, 1.1, 101,4°) | (-0.192, 1.1, 102.5°) |
| $j(X)$[d] (Å$^{-1}$) | $j$(O3): -0.0109 | $j$(O4): -0.0062 |
| ($\Delta h(X)$[e] Å, $l_n'/l_n$[f], CuXCu[g]) | (-0.049, 1.2, 96,2°) | (-0.027, 1.1, 94,9°) |
| $J$n/$J$max | $J2/J5$ = 0.59 | $J2/J5$ = 0.63 |
| Bond | Cu2-Cu2 | Cu3-Cu3 |
| d(Cu-Cu) (Å) | 3.120 | 3.064 |
| $J3$[c] (Å$^{-1}$) | $J3$ = -0.0395 (AFM) | $J3$ = -0.0365 (AFM) |
| $j(X)$[d] (Å$^{-1}$) | $j$(O1): -0.0395 | $j$(O1): -0.0365 |
| ($\Delta h(X)$[e] Å, $l_n'/l_n$[f], CuXCu[g]) | (-0.192, 1.0, 104,5°) | (-0.171, 1.0, 102,5°) |
| $J$n/$J$max | $J3/J5$ = 0.50 | $J3/J5$ = 0.47 |
| Bond | Cu2-Cu3 | Cu2-Cu3 |
| d(Cu-Cu) (Å) | 3.133 | 3.105 |
| $J4$[c] (Å$^{-1}$) | $J4$ = -0.0109 (AFM)[h] | $J4$ = -0.0060 (AFM) |
| $j(X)$[d] (Å$^{-1}$) | $j$(O4): -0.0109 | $j$(O3): -0.0060 |
| ($\Delta h(X)$[e] Å, $l_n'/l_n$[f], CuXCu[g]) | (-0.053, 1.2, 98.3°) | (-0.028, 1.2, 96.6°) |
| $J$n/$J$max | $J4/J5$ = 0.14 | $J4/J5$ = 0.08 |

| Bond | Cu2-Cu2 | Cu3-Cu3 |
|---|---|---|
| d(Cu-Cu) (Å) | 3.292 | 3.206 |
| $J5^{(c)}$ (Å$^{-1}$) | $J5 = -0.0789$ (AFM) | $J5 = -0.0777$ (AFM) |
| $j(X)^d$ (Å$^{-1}$) | $j$(O6): -0.0789 | $j$(O7): -0.0777 |
| ($\Delta h(X)^e$ Å, $l_n'/l_n^f$, CuXCu$^g$) | (-0.428, 1.0, 118.8°) | (-0.428, 1.0, 118,8°) |
| $J$n/$J$max | $J5/J5 = 1$ | $J5/J5 = 1$ |
| Bond | Cu2-Cu3 | Cu2-Cu3 |
| d(Cu-Cu) (Å) | 3.300 | 3.255 |
| $J6^{(c)}$ (Å$^{-1}$) | $J6 = -0.0635$ (AFM) | $J6 = -0.0654$ (AFM) |
| $j(X)^d$ (Å$^{-1}$) | $j$(O6): -0.0635 | $j$(O7): -0.0654 |
| ($\Delta h(X)^e$ Å, $l_n'/l_n^f$, CuXCu$^g$) | (-0.345, 1.07, 114,8°) | (-0.346, 1.07, 114,1°) |
| $J$n/$J$max | $J6/J5 = 0.80$ | $J6/J5 = 0.84$ |
| Bond | Cu3-Cu3 | Cu2-Cu2 |
| d(Cu-Cu) (Å) | 4.189 | 4.166 |
| $J7^{(c)}$ (Å$^{-1}$) | $J7 = 0.0284$ (FM) | $J7 = 0.0252$ (FM) |
| $j(X)^d$ (Å$^{-1}$) | $j$(O4): 0.0098x2 | $j$(O3): 0.0094x2 |
| ($\Delta h(X)^e$ Å, $l_n'/l_n^f$, CuXCu$^g$) | (0.450, 2.6, 90.6°) | (0.380, 2.3, 93.6°) |
| $j(X)^d$ (Å$^{-1}$) | $j$(O2): 0.0039x2 | $j$(O2): 0.0032x2 |
| ($\Delta h(X)^e$ Å, $l_n'/l_n^f$, CuXCu$^g$) | (0.180, 2.6, 98.5°) | (0.140, 2.5, 100.3°) |
| $J$n/$J$max | $J7/J5 = -0.36$ | $J7/J5 = -0.32$ |
| Bond | Cu2-Cu2 | Cu3-Cu3 |
| d(Cu-Cu) (Å) | 4.884 | 4.807 |
| $J8^{(c)}$ (Å$^{-1}$) | $J8 = 0.0050$ (FM) | $J8 = 0.0046$ (FM) |
| $j(X)^d$ (Å$^{-1}$) | $j$(O4): 0.0025x2 | $j$(O3): 0.0023x2 |
| ($\Delta h(X)^e$ Å, $l_n'/l_n^f$, CuXCu$^g$) | (0.200, 3.3, 102.1°) | (0.161, 3.0, 104.3°) |
| $J$n/$J$max | $J8/J5 = -0.06$ | $J8/J5 = -0.06$ |
| Bond | Cu1-Cu2 | Cu1-Cu3 |
| d(Cu-Cu) (Å) | 5.362 | 5.292 |
| $J9^{(c)}$ (Å$^{-1}$) | $J9^h = -0.0453$ (AFM) | $J9^h = -0.0473$ (AFM) |
| $j(X)^d$ (Å$^{-1}$) | $j$(O5): -0.0551 | $j$(O5): -0.0551 |
| ($\Delta h(X)^e$ Å, $l_n'/l_n^f$, CuXCu$^g$) | (-0.0670, 1.8, 147.1°) | (-0.0670, 1.8, 147.1°) |
| $J$n/$J$max | $J9/J5 = 0.57$ | $J9/J5 = 0.64$ |
| Bond | Cu2-Cu3 | Cu2-Cu3 |
| d(Cu-Cu) (Å) | 5.561 | 5.454 |
| $J10^{(c)}$ (Å$^{-1}$) | $J10^h = -0.0440$ (AFM) | $J10^h = -0.0516$ (AFM) |
| $j(X)^d$ (Å$^{-1}$) | $j$(O4): -0.0432 | $j$(O3): -0.0514 |
| ($\Delta h(X)^e$ Å, $l_n'/l_n^f$, CuXCu$^g$) | (-0.574, 1.7, 144.6°) | (-0.676, 1.7, 148.5°) |
| $J$n/$J$max | $J10/J5 = 0.56$ | $J10/J5 = 0.66$ |
| Bond | Cu1-Cu1 | Cu1-Cu1 |
| d(Cu-Cu) (Å) | 8.009 | 8.002 |
| $J11^{(c)}$ (Å$^{-1}$) | $J11^h = -0.0253$ (FM) | $J11^h = -0.0218$ (FM) |
| Bond | Cu2-Cu2 | Cu2-Cu2 |
| d(Cu-Cu) (Å) | 8.129 | 8.088 |
| $J12^{(c)}$ (Å$^{-1}$) | $J12^h = 0.0266$ (FM) | $J12^h = 0.0289$ (FM) |
| Bond | Cu1-Cu1 | Cu1-Cu1 |
| d(Cu-Cu) (Å) | 6.412 | 6.270 |
| $J_b^{1-1\,(c)}$ (Å$^{-1}$) | $J_b^{1-1} = 0.0086$ (FM) | $J_b^{1-1} = -0.0027$ (AFM) |
| $j(X)^d$ (Å$^{-1}$) | $j$(O3): 0.0184 | $j$(O4): 0.0073 |
| ($\Delta h(X)^e$ Å, $l_n'/l_n^f$, CuXCu$^g$) | (0.376, 1.0, 122.04°) | (0.144, 1.0, 127.55°) |
| $j(X)^d$ (Å$^{-1}$) | $j$(O5): -0.0049x2 | $j$(O6): -0.0050x2 |
| ($\Delta h(X)^e$ Å, $l_n'/l_n^f$, CuXCu$^g$) | (-0.498, 2.5, 142.7°) | (-0.457, 2.3, 141.3°) |
| $J$n/$J$max | $J_b^{1-1}/J5 = -0.11$ | $J_b^{1-1}/J5 = 0.03$ |
| Bond | Cu2-Cu2 | Cu3-Cu3 |
| d(Cu-Cu) (Å) | 6.412 | 6.270 |
| $J_b^{2-2\,(c)}$ (Å$^{-1}$) | $J_b^{2-2} = -0.0329$ (AFM) | $J_b^{3-3} = -0.0339$ (AFM) |
| $j(X)^d$ (Å$^{-1}$) | $j$(Cu2): -0.0278 | $j$(Cu3): -0.0290 |
| ($\Delta h$(X)$^e$ Å, $l_n$'/$l_n^f$, CuXCu$^g$) | (-0.570, 1.06, 180°) | (-0.570, 1.05, 180°) |
| $j(X)^d$ (Å$^{-1}$) | $j$(O6): -0.0036 | $j$(O7): -0.0035 |
| ($\Delta h$(X)$^e$ Å, $l_n$'/$l_n^f$, CuXCu$^g$) | (-0.428, 2.9, 137.9°) | (-0.399, 2.9, 135.9°) |
| $j(X)^d$ (Å$^{-1}$) | $j$(O1): -0.0015 | $j$(O5): -0.0014 |
| ($\Delta h$(X)$^e$ Å, $l_n$'/$l_n^f$, CuXCu$^g$) | (-0.192, 3.1, 128.3°) | (-0.171, 3.1, 126.7°) |
| $J$n/$J$max | $J_b^{2-2}/J5 = 0.42$ | $J_b^{3-3}/J5 = 0.42$ |
| Bond | Cu3-Cu3 | Cu2-Cu2 |
| d(Cu-Cu) (Å) | 6.412 | 6.270 |
| $J_b^{3-3\,(c)}$ (Å$^{-1}$) | $J_b^{3-3} = 0204$ (FM) | $J_b^{2-2} = 0.0066$ (FM) |
| $j(X)^d$ (Å$^{-1}$) | $j$(O7): 0.0241 | $j$(O7): 0.0241 |
| ($\Delta h$(X)$^e$ Å, $l_n$'/$l_n^f$, CuXCu$^g$) | (0.495, 1.0, 118.8°) | (0.495, 1.0, 118.8°) |
| $j(X)^d$ (Å$^{-1}$) | $j$(As): 0.0080 | $j$(P): 0.0105 |
| ($\Delta h$(X)$^e$ Å, $l_n$'/$l_n^f$, CuXCu$^g$) | (0.165, 1.0, 162.3°) | (0.206, 1.0, 166.3°) |
| $j(X)^d$ (Å$^{-1}$) | $j$(O2): -0.0091 | $j$(O2): -0.0109 |
| ($\Delta h$(X)$^e$ Å, $l_n$'/$l_n^f$, CuXCu$^g$) | (-0.187, 1.0, 138.6°) | (-0.215, 1.0, 138.6°) |
| $j(X)^d$ (Å$^{-1}$) | $j$(O4): -0.0013x2 | $j$(O3): -0.0028x2 |
| ($\Delta h$(X)$^e$ Å, $l_n$'/$l_n^f$, CuXCu$^g$) | (-0.136, 2.6, 129.3°) | (-0.263, 2.4, 134.04°) |
| $J$n/$J$max | $J_b^{3-3}/J5 = -0.26$ | $J_b^{2-2}/J5 = -0.08$ |
| **Between-ribbon couplings** | | |
| Bond | Cu1 - Cu1 | Cu1 - Cu1 |
| d(Cu-Cu) (Å) | 4.251 | 4.057 |
| $J11^{(c)}$ (Å$^{-1}$) | $J11 = 0.0170$ (FM) | $J11 = 0.0208$ (FM) |
| $j(X)^d$ (Å$^{-1}$) | $j$(O3): 0.0085x2 | $j$(O4): 0.0104x2 |
| ($\Delta h$(X)$^e$ Å, $l_n$'/$l_n^f$, CuXCu$^g$) | (0.447, 2.9, 90.2°) | (0.460, 2.7, 88.4°) |
| $J$n/$J$max | $J11/J5 = -0.21$ | $J11/J5 = -0.27$ |
| Bond | Cu1-Cu3 | Cu1-Cu2 |
| d(Cu-Cu) (Å) | 5.914 | 5.757 |
| $J12^{(c)}$ (Å$^{-1}$) | $J12 = -0.0411$ (AFM) | $J12 = -0.0345$ (AFM) |
| $j(X)^d$ (Å$^{-1}$) | $j$(O4): -0.0375 | $j$(O3): -0.0305 |
| ($\Delta h$(X)$^e$ Å, $l_n$'/$l_n^f$, CuXCu$^g$) | (-0.531, 1.95, 144.03°) | (-0.407, 1.98, 138.3°) |
| $j(X)^d$ (Å$^{-1}$) | $j$(O5): -0.0036 | $j$(O6): -0.0040 |
| ($\Delta h$(X)$^e$ Å, $l_n$'/$l_n^f$, CuXCu$^g$) | (-0.304, 2.4, 132.9°) | (-0.297, 2.2, 132.8°) |
| $J$n/$J$max | $J12/J5 = 0.52$ | $J12/J5 = 0.44$ |
| Bond | Cu1-Cu2 | Cu1-Cu3 |
| d(Cu-Cu) (Å) | 6.212 | 6.047 |
| $J13^{(c)}$ (Å$^{-1}$) | $J13 = -0.0173$ (AFM) | $J13 = -0.0209$ (AFM) |
| $j(X)^d$ (Å$^{-1}$) | $j$(O1): -0.0129 | $j$(O5): -0.0160 |
| ($\Delta h$(X)$^e$ Å, $l_n$'/$l_n^f$, CuXCu$^g$) | (-1.089, 2.19, 166.64°) | (-1.224, 2.1, 172.4°) |
| $j(X)^d$ (Å$^{-1}$) | $j$(O5): -0.0044 | $j$(O6): -0.0049 |
| ($\Delta h$(X)$^e$ Å, $l_n$'/$l_n^f$, CuXCu$^g$) | (-0.418, 2.45, 138.45°) | (-0.406, 2.3, 138.5°) |
| $J$n/$J$max | $J13/J5 = 0.22$ | $J13/J5 = 0.27$ |
| Bond | Cu1-Cu3 | Cu1-Cu2 |
| d(Cu-Cu) (Å) | 6.244 | 6.058 |
| $J14^{(c)}$ (Å$^{-1}$) | $J14 = -0.0199$ (AFM) | $J14 = -0.0241$ (AFM) |
| $j(X)^d$ (Å$^{-1}$) | $j$(O2): -0.0121 | $j$(O2): -0.0145 |
| ($\Delta h$(X)$^e$ Å, $l_n$'/$l_n^f$, CuXCu$^g$) | (-1.063, 2.2, 165.62°) | (-1.142, 2.1, 168.8°) |
| $j(X)^d$ (Å$^{-1}$) | $j$(O7): -0.078 | $j$(O1): -0.096 |
| ($\Delta h$(X)$^e$ Å, $l_n$'/$l_n^f$, CuXCu$^g$) | (-0.785, 2.6, 152.74°) | (-0.863, 2.4, 155.9°) |
| $J$n/$J$max | $J14/J5 = 0.25$ | $J14/J5 = 0.31$ |
| Bond | Cu1-Cu2 | Cu1-Cu3 |
| d(Cu-Cu) (Å) | 6.262 | 6.064 |
| $J15^{(c)}$ (Å$^{-1}$) | $J15 = -0.0210$ (AFM) | $J15 = -0.0245$ (AFM) |
| $j(X)^d$ (Å$^{-1}$) | $j$(O4): -0.0124 | $j$(O3): -0.0155 |
| ($\Delta h$(X)$^e$ Å, $l_n$'/$l_n^f$, CuXCu$^g$) | (-1.090, 2.2, 166.8°) | (-1.195, 2.1, 171.2°) |
| $j(X)^d$ (Å$^{-1}$) | $j$(O7): -0.086 | $j$(O1): -0.090 |
| ($\Delta h(X)^e$ Å, $l_n'$/$l_n^f$, CuXCu$^g$) | (-0.865, 2.5, 156.5°) | (-0.818, 2.5, 153.9°) |
| $J$n/$J$max | $J15/J5 = 0.27$ | $J15/J5 = 0.32$ |
| Bond | Cu2-Cu3 | Cu2-Cu3 |
| d(Cu-Cu) (Å) | 6.299 | 6.106 |
| $J16^{(c)}$ (Å$^{-1}$) | $J16 = -0.0188$ (AFM) | $J16 = -0.0219$ (AFM) |
| $j(X)^d$ (Å$^{-1}$) | $j$(O3): -0.0088 | $j$(O4): -0.0101 |
| ($\Delta h(X)^e$ Å, $l_n'$/$l_n^f$, CuXCu$^g$) | (-0.847, 2.4, 156.2°) | (-0.870, 2.3, 156.9°) |
| $j(X)^d$ (Å$^{-1}$) | $j$(O5): -0.100 | $j$(O6): -0.118 |
| ($\Delta h(X)^e$ Å, $l_n'$/$l_n^f$, CuXCu$^g$) | (-0.912, 2.3, 159.4°) | (-0.942, 2.1, 160.5°) |
| $J$n/$J$max | $J16/J5 = 0.24$ | $J16/J5 = 0.28$ |
| Bond | Cu2-Cu3 | Cu2-Cu3 |
| d(Cu-Cu) (Å) | 6.565 | 6.345 |
| $J17^{(c)}$ (Å$^{-1}$) | $J17 = 0.0189$ (FM) | $J17 = 0.0196$ (FM) |
| $j(X)^d$ (Å$^{-1}$) | $j$(O7): 0.0230 | $j$(O1): 0.0236 |
| ($\Delta h(X)^e$ Å, $l_n'$/$l_n^f$, CuXCu$^g$) | (0.495, 1.05, 120.0°) | (0.473, 1.1, 118.8°) |
| $j(X)^d$ (Å$^{-1}$) | $j$(O4): -0.0041 | $j$(O3): -0.0040 |
| ($\Delta h(X)^e$ Å, $l_n'$/$l_n^f$, CuXCu$^g$) | (-0.416, 2.4, 141.2°) | (-0.375, 2.3, 138.8°) |
| $J$n/$J$max | $J17/J5 = -0.24$ | $J17/J5 = -0.25$ |
| Bond | Cu1-Cu3 | Cu1-Cu2 |
| d(Cu-Cu) (Å) | 6.640 | 6.494 |
| $J18^{(c)}$ (Å$^{-1}$) | $J18 = 0.0329$ (FM) | $J18 = 0.0283$ (FM) |
| $j(X)^d$ (Å$^{-1}$) | $j$(O4): 0.0184x2 | $j$(O3): 0.0161x2 |
| ($\Delta h(X)^e$ Å, $l_n'$/$l_n^f$, CuXCu$^g$) | (0.402, 1.2, 122.8°) | (0.334, 1.2, 123.5°) |
| $j(X)^d$ (Å$^{-1}$) | $j$(O6): -0.0039 | $j$(O7): -0.0039 |
| ($\Delta h(X)^e$ Å, $l_n'$/$l_n^f$, CuXCu$^g$) | (-0.468, 2.7, 141.5°) | (-0.444, 2.7, 139.9°) |
| $J$n/$J$max | $J18/J5 = -0.42$ | $J18/J5 = -0.36$ |
| Bond | Cu3-Cu3 | Cu2-Cu2 |
| d(Cu-Cu) (Å) | 6.838 | 6.600 |
| $J19^{(c)}$ (Å$^{-1}$) | $J19^h = -0.0524$ (AFM) | $J19^h = -0.0586$ (AFM) |
| $j(X)^d$ (Å$^{-1}$) | $j$(O1): -0.0575 | $j$(O5): -0.0640 |
| ($\Delta h(X)^e$ Å, $l_n'$/$l_n^f$, CuXCu$^g$) | (-1.340, 1.1, 178.0°) | (-1.390, 1.1, 179.7°) |
| $J$n/$J$max | $J19/J5 = 0.73$ | $J19/J5 = 0.67$ |
| Bond | Cu1-Cu2 | Cu1-Cu3 |
| d(Cu-Cu) (Å) | 6.924 | 6.689 |
| $J20^{(c)}$ (Å$^{-1}$) | $J20^h = 0.0169$ (FM) | $J20^h = 0.0224$ (FM) |
| $j(X)^d$ (Å$^{-1}$) | $j$(Cu1): 0.0151 | $j$(Cu1): 0.0192 |
| ($\Delta h(X)^e$ Å, $l_n'$/$l_n^f$, CuXCu$^g$) | (0.335, 1.5, 147.2°) | (0.405, 1.4, 144.1°) |
| $J$n/$J$max | $J20/J5 = -0.21$ | $J20/J5 = -0.21$ |
| Bond | Cu2-Cu3 | Cu2-Cu3 |
| d(Cu-Cu) (Å) | 7.029 | 6.871 |
| $J21^{(c)}$ (Å$^{-1}$) | $J21^h = -0.0298$ (AFM) | $J21^h = -0.0363$ (AFM) |
| $j(X)^d$ (Å$^{-1}$) | $j$(O4): -0.0312 | $j$(O3): -0.0379 |
| ($\Delta h(X)^e$ Å, $l_n'$/$l_n^f$, CuXCu$^g$) | (-0.0770, 1.0, 159.7°) | (-0.0893, 1.0, 163.2°) |
| $J$n/$J$max | $J21/J5 = 0.38$ | $J21/J5 = 0.47$ |
| Bond | Cu1-Cu3 | Cu1-Cu2 |
| d(Cu-Cu) (Å) | 7.091 | 6.956 |
| $J22^{(c)}$ (Å$^{-1}$) | $J22^h = 0.0133$ (FM) | $J22^h = 0.0056$ (FM) |
| $j(X)^d$ (Å$^{-1}$) | $j$(O7): 0.0097 | $j$(O2): 0.0044 |
| ($\Delta h(X)^e$ Å, $l_n'$/$l_n^f$, CuXCu$^g$) | (0.243, 1.1, 130.2°) | (0.102, 1.3, 132.6°) |
| $j(X)^d$ (Å$^{-1}$) | $j$(O2): 0.0021 | $j$(O3): 0.0017 |
| ($\Delta h(X)^e$ Å, $l_n'$/$l_n^f$, CuXCu$^g$) | (0.051, 1.3, 134.9°) | (0.384, 4.6, 107.5°) |
| $J$n/$J$max | $J22/J5 = -0.17$ | $J22/J5 = -0.07$ |
| Bond | Cu3-Cu3 | Cu2-Cu2 |
| d(Cu-Cu) (Å) | 7.825 | 7.589 |

| | $J22^{(c)}$ (Å$^{-1}$) | $J23^h$ = -0.0170 (FM) | $J23^h$ = 0.0195 (FM) |
|---|---|---|---|
| | $j(X)^d$ (Å$^{-1}$) | $j$(Cu2): 0.0077 | $j$(Cu3): 0.0070 |
| | $(\Delta h(X)^e$ Å, $l_n'/l_n^f$, CuXCu$^g$) | (0.222, 1.3, 156.4 °) | (0.191, 1.4, 156.7 °) |
| | $j(X)^d$ (Å$^{-1}$) | $j$(O1): 0.0167 | $j$(O5): 0.0188 |
| | $(\Delta h(X)^e$ Å, $l_n'/l_n^f$, CuXCu$^g$) | (0.461, 1.6, 127.2 °) | (0.478, 1.7, 125.0°) |
| | $J$n/$J$max | $J23/J5$ = -0.22 | $J23/J5$ = -0.25 |

[a]XDS: X-ray diffraction from a single crystal.
[b]The refinement converged to the residual factor ($R$) values.
[c]$J$n in Å$^{-1}$: the magnetic couplings ($J$n < 0, AFM; $J$n > 0, FM).
[d]$j$(X): contributions of the intermediate ion X to the AFM ($j$(X) < 0) and FM ($j$(X) > 0) components of the coupling $J$n.
[e]$\Delta h$(X): the degree of overlapping of the local space between magnetic ions by the intermediate ion X.
[f]$l_n'/l_n$: the asymmetry of the position of the intermediate ion X relative to the middle of the Cu$_i$–Cu$_j$ bond line.
[g]Cu$_i$XCu$_j$: bonding angle.
[h]Small j(X) contributions are not shown.

**Table 2.** Crystallographic characteristics and parameters of magnetic couplings ($J$n) calculated on the basis of structural data and respective distances between magnetic ions Cu$^{2+}$ in the copper(II) oxide arsenate ericlaxmanite Cu$_4$O(AsO$_4$)$_2$ and phosphate Cu$_4$O(PO$_4$)$_2$

| Crystallographic and magnetic parameters | Cu$_4$O(AsO$_4$)$_2$[35] Min Name: Ericlaxmanite (Data for ICSD-404850) Space group $P$ -1 (N2) $a$ = 6.415, $b$ = 7.655, $c$ = 8.224 Å $\alpha$ = 98.52, $\beta$ = 112.39, $\gamma$ = 98.38º, Z = 2 Method$^{(a)}$: XDS (293 K); $R$-value$^{(b)}$ = 0.038 | Cu$_4$O(PO$_4$)$_2$[14] - (Data for ICSD-1666) Space group $P$ -1 (N2) $a$ =7.528, $b$ = 8.090, $c$ = 6.272 Å $\alpha$ = 113.68, $\beta$ = 81.56, $\gamma$ = 105.77º, Z = 2 Method$^{(a)}$: XDS (293 K); $R$-value$^{(b)}$ = 0.038 |
|---|---|---|
| d(Cu-X) (Å) | **Cu1:** tetragonal pyramid<br>Cu1-O3 = 1.926<br>-O6 = 1.933<br>-O9 = 1.939<br>-O8 = 1.959<br>-O8 = 2.648<br><br>**Cu2:** [4+2] octahedron<br>Cu2-O3 = 1.903<br>-O1 = 1.962<br>-O7 = 1.987<br>-O5 = 2.045<br>-O8 = 2.316<br>-O4 = 2.779<br>**Cu3:** trigonal bipyramid<br>Cu3-O3 = 1.911<br>-O4 = 1.941<br>-O1 = 1.957<br>-O9 = 2.015<br>-O7 = 2.464<br><br>**Cu4:** [4+2] octahedron<br>Cu4-O3 = 1.909<br>-O2 = 1.966<br>-O5 = 1.973<br>-O2 = 2.011<br>-O6 = 2.457<br>-O4 = 2.462 | **Cu2:** tetragonal pyramid (square)<br>Cu2-O6 = 1.916<br>-O1 = 1.921<br>-O9 = 1.934<br>-O7 = 1.963<br>-O2 = 2.951<br>-O7 = 2.978<br><br>**Cu3:** [4+2] octahedron<br>Cu3-O1 = 1.901<br>-O2 = 1.947<br>-O8 = 2.028<br>-O4 = 2.052<br>-O7 = 2.230<br>-O5 = 2.794<br>**Cu1**: trigonal bipyramid<br>Cu1-O1 = 1.898<br>-O5 = 1.948<br>-O2 = 1.950<br>-O9 = 2.085<br>-O8 = 2.289<br>-O4 = 2.997<br><br>**Cu4:** [4+2] octahedron<br>Cu4-O1 = 1.914<br>-O4 = 1.973<br>-O3 = 1.981<br>-O3 = 2.001<br>-O6 = 2.435<br>-O5 = 2.521 |

| *Complicated ribbon* | | |
|---|---|---|
| Bond | Cu1-Cu2 | Cu2-Cu3 |
| d(Cu-Cu) (Å) | 2.882 | 2.856 |
| $J1^{(c)}$ (Å$^{-1}$) | $J1$ = -0.0335 (AFM) | $J1$ = -0.0318 (AFM) |
| $j(X)^d$ (Å$^{-1}$) | $j$(O3): - 0.0335 | $j$(O1): - 0.0318 |
| ($\Delta h(X)^e$ Å, $l_n'/l_n^f$, CuXCu$^g$) | (-0.139, 1.0, 97.61°) | (-0.130, 1.0, 96.70°) |
| $Jn/J$max | $J1/J10$ = 0.29 | $J1/J10$ = 0.27 |
| Bond | Cu1-Cu3 | Cu2-Cu1 |
| d(Cu-Cu) (Å) | 2.897 | 2.871 |
| $J2^{(c)}$ (Å$^{-1}$) | $J2$ = -0.0469 (AFM) | $J2$ = -0.0330 (AFM) |
| $j(X)^d$ (Å$^{-1}$) | $j$(O3): -0.0338 | $j$(O1): -0.0340 |
| ($\Delta h(X)^e$ Å, $l_n'/l_n^f$, CuXCu$^g$) | (-0.142, 1.0, 98,1°) | (-0.140, 1.0, 97,4°) |
| $j(X)^d$ (Å$^{-1}$) | $j$(O9): -0.0132 | $j$(O9): 0.0010 |
| ($\Delta h(X)^e$ Å, $l_n'/l_n^f$, CuXCu$^g$) | (-0.055, 1.1, 94,2°) | (0.004, 1.2, 91,1°) |
| $Jn/J$max | $J2/J10$ = 0.41 | $J2/J10$ = 0.28 |
| Bond | Cu2-Cu4 | Cu3-Cu4 |
| d(Cu-Cu) (Å) | 2.927 | 2.931 |
| $J3^{(c)}$ (Å$^{-1}$) | $J3$ = -0.0474 (AFM) | $J3$ = -0.0467 (AFM) |
| $j(X)^d$ (Å$^{-1}$) | $j$(O3): -0.0417 | $j$(O1): -0.0417 |
| ($\Delta h(X)^e$ Å, $l_n'/l_n^f$, CuXCu$^g$) | (-0.179, 1.0, 100.3°) | (-0.179, 1.0, 100.4°) |
| $j(X)^d$ (Å$^{-1}$) | $j$(O5): -0.0057 | $j$(O4): -0.0049 |
| ($\Delta h(X)^e$ Å, $l_n'/l_n^f$, CuXCu$^g$) | (-0.024, 1.1, 93,5°) | (-0.021, 1.1, 93,5°) |
| $Jn/J$max | $J3/J10$ = 0.41 | $J3/J10$ = 0.39 |
| Bond (*between-ribbon*) | Cu4-Cu4 | Cu4-Cu4 |
| d(Cu-Cu) (Å) | 2.938 | 2.913 |
| $J4^{(c)}$ (Å$^{-1}$) | $J4$ = -0.0280(AFM) | $J4$ = -0.0202(AFM) |
| $j(X)^d$ (Å$^{-1}$) | $j$(O2): -0.0140x2 | $j$(O3): -0.0101x2 |
| ($\Delta h(X)^e$ Å, $l_n'/l_n^f$, CuXCu$^g$) | (-0.060, 1.0, 95.3°) | (-0.043, 1.0, 94.03°) |
| $Jn/J$max | $J4/J10$ = 0.24 | $J4/J10$ = 0.18 |
| Bond | Cu3-Cu4 | Cu1-Cu4 |
| d(Cu-Cu) (Å) | 3.098 | 3.105 |
| $J5^{(c)}$ (Å$^{-1}$) | $J5$ = -0.0588 (AFM) | $J5$ = -0.0611 (AFM) |
| $j(X)^d$ (Å$^{-1}$) | $j$(O3): -0.0588 | $j$(O1): -0.0611 |
| ($\Delta h(X)^e$ Å, $l_n'/l_n^f$, CuXCu$^g$) | (-0.282, 1.0, 108.4°) | (-0.294, 1.0, 108.4°) |
| $Jn/J$max | $J5/J10$ = 0.51 | $J5/J10$ = 0.51 |
| Bond | Cu2-Cu3 | Cu1-Cu3 |
| d(Cu-Cu) (Å) | 3.122 | 3.041 |
| $J6^{(c)}$ (Å$^{-1}$) | $J6$ = -0.0624 (AFM) | $J6$ = -0.0564 (AFM) |
| $j(X)^d$ (Å$^{-1}$) | $j$(O3): -0.0624 | $j$(O1): -0.0564 |
| ($\Delta h(X)^e$ Å, $l_n'/l_n^f$, CuXCu$^g$) | (-0.304, 1.0, 109,9°) | (-0.261, 1.0, 106,3°) |
| $Jn/J$max | $J6/J10$ = 0.54 | $J6/J10$ = 0.47 |
| Bond | Cu1-Cu1 | Cu2-Cu2 |
| d(Cu-Cu) (Å) | 3.288 | 3.457 |
| $J7^{(c)}$ (Å$^{-1}$) | $J7$ = 0 | $J7$ = 0 |
| Bond | Cu2-Cu3 | Cu1-Cu3 |
| d(Cu-Cu) (Å) | 3.311 | 3.239 |
| $J8^{(c)}$ (Å$^{-1}$) | $J8$ = -0.0642 (AFM) | $J8$ = -0.0604 (AFM) |
| $j(X)^d$ (Å$^{-1}$) | $j$(O1): -0.0642 | $j$(O2): -0.0604 |
| ($\Delta h(X)^e$ Å, $l_n'/l_n^f$, CuXCu$^g$) | (-0.352, 1.0, 115.3°) | (-0.317, 1.0, 115.3°) |
| $Jn/J$max | $J8/J10$ = 0.56 | $J8/J10$ = 0.51 |
| Bond | Cu1-Cu4 | Cu2-Cu4 |
| d(Cu-Cu) (Å) | 3.595 | 3.450 |
| $J9^{(c)}$ (Å$^{-1}$) | $J9$ = 0 | $J9$ = 0 |
| Bond | Cu1-Cu4 | Cu2-Cu4 |
| d(Cu-Cu) (Å) | 3.609 | 3.632 |
| $J10^{(c)}$ (Å$^{-1}$) | $J10$ = -0.1154 (AFM) | $J10$ = -0.1190 (AFM) |
| $j(X)^d$ (Å$^{-1}$) | $j$(O3): -0.01154 | $j$(O7): -0.01190 |
| $(\Delta h(X)^e$ Å, $l_n'/l_n^f$, CuXCu$^g$) | (-0.751, 1.0, 140.5°) | (-0.785, 1.0, 142.6°) |
| $J$n/$J$max | $J10/J10$ = 1 | $J10/J10$ = 1 |
| Bond | Cu2-Cu3 | Cu1-Cu3 |
| d(Cu-Cu) (Å) | 3.990 | 3.834 |
| $J11^{(c)}$ (Å$^{-1}$) | [i]$J11$ = 0.0062 (FM) | [i]$J11$ = 0. |
| $j(X)^d$ (Å$^{-1}$) | $j$(O5): 0.0062 (FM) | |
| $(\Delta h(X)^e$ Å, $l_n'/l_n^f$, CuXCu$^g$) | (0.216, 2.2, 97.2°) | |
| $J$n/$J$max | $J11/J10$ = -0.04 | |
| Bond | Cu1-Cu2 | Cu2-Cu3 |
| d(Cu-Cu) (Å) | 4.124 | 4.119 |
| $J12^{(c)}$ (Å$^{-1}$) | $J12$ = 0.0041 (FM) | $J12$ = 0.0040 (FM) |
| $j(X)^d$ (Å$^{-1}$) | $j$(O3): 0.0041 | $j$(O1): 0.0040 |
| $(\Delta h(X)^e$ Å, $l_n'/l_n^f$, CuXCu$^g$) | (0.220, 3.1, 94.3°) | (0.207, 3.1, 94.9°) |
| $J$n/$J$max | $J12/J10$ = -0.05 | $J12/J10$ = -0.03 |
| Bond | Cu1-Cu2 | Cu2-Cu3 |
| d(Cu-Cu) (Å) | 4.355 | 4.141 |
| $J13^{(c)}$ (Å$^{-1}$) | [i]$J13$ = -0.0021 ↔ -0.0105(AFM) | [i]$J13$ =0 |
| $j(X)^d$ (Å$^{-1}$) | $j$(O1): -0.0021↔ -0.0105(AFM) | |
| $(\Delta h(X)^e$ Å, $l_n'/l_n^f$, CuXCu$^g$) | (-0.079, [i]2.0, 113.2°) | |
| $J$n/$J$max | $J13/J10$ = 0.02↔0.09 | |
| Bond | Cu2-Cu2 | Cu3-Cu3 |
| d(Cu-Cu) (Å) | 4.439 | 4.270 |
| $J14^{(c)}$ (Å$^{-1}$) | $J14$ = 0.0084 (FM) | $J14$ = 0.0120 FM |
| $j(X)^d$ (Å$^{-1}$) | $j$(O7): 0.0042 (FM)x2 | $j$(O7): 0.0060 (FM)x2 |
| $(\Delta h(X)^e$ Å, $l_n'/l_n^f$, CuXCu$^g$) | (0.248, 3.0, 97.6°) | (0.336, 3.1, 92.8°) |
| $J$n/$J$max | $J14/J10$ = -0.07 | $J14/J10$ = -0.10 |
| Bond | Cu3-Cu4 | Cu1-Cu4 |
| d(Cu-Cu) (Å) | 4.450 | 4.245 |
| $J15^{(c)}$ (Å$^{-1}$) | [i]$J15$ = -0.0185 (AFM) | [i]$J15$ = 0. |
| $j(X)^d$ (Å$^{-1}$) | $j$(O5): -0.0185 | |
| $(\Delta h(X)^e$ Å, $l_n'/l_n^f$, CuXCu$^g$) | (-0.151, [i]1.9, 117.6°) | |
| $J$n/$J$max | $J15/J10$= 0.16 | |
| Bond | Cu1-Cu3 | Cu2-Cu1 |
| d(Cu-Cu) (Å) | 4.522 | 4.250 |
| $J16^{(c)}$ (Å$^{-1}$) | [i]$J16$ = -0.0224 (AFM) | [i]$J16$ = -0.0043 (FM) |
| $j(X)^d$ (Å$^{-1}$) | $j$(O1): -0.0257 | $j$(O9): 0.0043 |
| $(\Delta h(X)^e$ Å, $l_n'/l_n^f$, CuXCu$^g$) | (-0.216, 1.9, 121.0°) | (0.254, 3.2, 94.2°) |
| $j(X)^d$ (Å$^{-1}$) | $j$(O9): 0.0033 | $J16/J10$ = -0.04 |
| $(\Delta h(X)^e$ Å, $l_n'/l_n^f$, CuXCu$^g$) | (0.217, 3.2, 98.4°) | - |
| $J$n/$J$max | $J16/J10$ = 0.19 | - |
| Bond | Cu3-Cu3 | Cu1-Cu1 |
| d(Cu-Cu) (Å) | 4.550 | 4.368 |
| $J17^{(c)}$ (Å$^{-1}$) | $J17$ = 0.0086 (FM) | $J17$ = 0.0126 FM |
| $j(X)^d$ (Å$^{-1}$) | $j$(O9): 0.0043x2 | $j$(O9): 0.0063x2 |
| $(\Delta h(X)^e$ Å, $l_n'/l_n^f$, CuXCu$^g$) | (0.272, 3.0, 97.9°) | (0.332, 2.8, 95.4°) |
| $J$n/$J$max | $J17/J10$ = -0.07 | $J17/J10$ = -0.11 |
| Bond | Cu1-Cu3 | Cu1-Cu2 |
| d(Cu-Cu) (Å) | 4.742 | 4.775 |
| $J18^{(c)}$ (Å$^{-1}$) | $J18$ = 0.0040 (FM) | $J18$ = 0.0047 (FM) |
| $j(X)^d$ (Å$^{-1}$) | $j$(O9): 0.0040 | $j$(O9): 0.0047 |
| $(\Delta h(X)^e$ Å, $l_n'/l_n^f$, CuXCu$^g$) | (0.317, 3.5, 96.6°) | (0.338, 3.1, 97.9°) |
| $J$n/$J$max | $J18/J10$ = -0.03 | $J18/J10$ = -0.04 |
| Bond | Cu1-Cu4 | Cu2-Cu4 |

| d(Cu-Cu) (Å) | 5.020 | 5.119 |
|---|---|---|
| $J19^{(c)}$ (Å$^{-1}$) | $J19$ = 0.0048 (FM) | $J19$ = -0.0048 (AFM) |
| $j(X)^d$ (Å$^{-1}$) | $j$(O5): 0.0026 | $j$(O3): -0.0048 |
| ($\Delta h(X)^e$ Å, $l_n$'/$l_n^f$, CuXCu$^g$) | (0.236, 3.6, 101.3°) | (-0.271, 2.1, 127.4°) |
| $j(X)^d$ (Å$^{-1}$) | $j$(O9): 0.0024 | $J19/J10$ = 0.04 |
| ($\Delta h(X)^e$ Å, $l_n$'/$l_n^f$, CuXCu$^g$) | (0.225, 3.7, 100.8°) | |
| $Jn/J$max | $J19/J10$ = -0.04 | |
| Bond | Cu2-Cu2 | Cu3-Cu3 |
| d(Cu-Cu) (Å) | 5.122 | 4.956 |
| $J20^{(c)}$ (Å$^{-1}$) | $J20$ = 0.0078 (FM) | $J20$ = 0.0074 (FM) |
| $j(X)^d$ (Å$^{-1}$) | $j$(O1): 0.0009x2 | $j$(O2): 0.0005x2 |
| ($\Delta h(X)^e$ Å, $l_n$'/$l_n^f$, CuXCu$^g$) | (0.068, 2.9, 110.5.°) | (0.034, 2.8, 111.1°) |
| $j(X)^d$ (Å$^{-1}$) | $j$(O5): 0.0030 x 2 | $j$(O4): 0.0032 x 2 |
| ($\Delta h(X)^e$ Å, $l_n$'/$l_n^f$, CuXCu$^g$) | (0.252, 3.2, 103.2.°) | (0.234, 3.0, 103.5.°) |
| $Jn/J$max | $J20/J10$ = -0.07 | $J20/J10$ = -0.06 |
| Bond | Cu3-Cu4 | Cu1-Cu4 |
| d(Cu-Cu) (Å) | 5.211 | 5.158 |
| $J21^{(c)}$ (Å$^{-1}$) | [i]$J21$ = -0.0423(AFM) ↔ -0.0021(AFM) | $J21$ = 0.0026 |
| $j(X)^d$ (Å$^{-1}$) | [i]$j$(O5): -0.0449(AFM) | [i]O5 -ax |
| ($\Delta h(X)^e$ Å, $l_n$'/$l_n^f$, CuXCu$^g$) | (-0.491, 2.0, 137.8°) | - |
| $j(X)^d$ (Å$^{-1}$) | $j$(O1): 0.0026 (FM) | $j$(O2): 0.0026 (FM) |
| ($\Delta h(X)^e$ Å, $l_n$'/$l_n^f$, CuXCu$^g$) | (0.311, 4.5, 97.2°) | (0.321, 4.6, 96.0°) |
| $Jn/J$max | $J21/J10$ = 0.37 | $J21/J10$ = -0.02 |
| Bond | Cu1-Cu3 | Cu1-Cu2 |
| d(Cu-Cu) (Å) | 5.244 | 5.222 |
| $J22^{(c)}$ (Å$^{-1}$) | $J22^h$ = -0.0625 (AFM) | $J22^h$ = -0.0629 (AFM) |
| $j(X)^d$ (Å$^{-1}$) | $j$(O8): -0.0674 | $j$(O7): -0.0684 |
| ($\Delta h(X)^e$ Å, $l_n$'/$l_n^f$, CuXCu$^g$) | (-0.782, 1.8, 151.3°) | (-0.792, 1.8, 151.7°) |
| $Jn/J$max | $J22/J10$ = 0.54 | $J22/J10$ = 0.53 |
| Bond | Cu1-Cu3 | Cu1-Cu2 |
| d(Cu-Cu) (Å) | 5.299 | 5.300 |
| $J23^{(c)}$ (Å$^{-1}$) | $J23^h$ = -0.0707 (AFM) | $J23^h$ = -0.0774 (AFM) |
| $j(X)^d$ (Å$^{-1}$) | $j$(O3): -0.0721 | $j$(O1): -0.0795 |
| ($\Delta h(X)^e$ Å, $l_n$'/$l_n^f$, CuXCu$^g$) | (-0.834, 1.9, 153.53°) | (-0.925, 1.9, 157.7°) |
| $Jn/J$max | $J23/J10$ = 0.61 | $J23/J10$ = 0.65 |
| Bond | Cu2-Cu4 | Cu3-Cu4 |
| d(Cu-Cu) (Å) | 5.407 | 5.246 |
| $J24^{(c)}$ (Å$^{-1}$) | $J24^h$ = 0.0017 (FM) | $J24^h$ = 0.0040 (FM) |
| $j(X)^d$ (Å$^{-1}$) | $j$(O2): 0.0007 | $j$(O2): 0.0024 |
| ($\Delta h(X)^e$ Å, $l_n$'/$l_n^f$, CuXCu$^g$) | (0.057, 2.9, 113.5°) | (0.299, 2.9, 97.7°) |
| $j(X)^d$ (Å$^{-1}$) | $j$(O7): 0.0010 | $j$(O8): 0.0016 |
| ($\Delta h(X)^e$ Å, $l_n$'/$l_n^f$, CuXCu$^g$) | (0.095, 3.1, 111.2°) | (0.133, 3.1, 111.2°) |
| $Jn/J$max | $J24/J10$ = -0.01 | $J24/J10$ = -0.03 |
| Bond | Cu4-Cu4 | Cu4-Cu4 |
| d(Cu-Cu) (Å) | 5.411 | 5.220 |
| $J25^{(c)}$ (Å$^{-1}$) | $J25^h$ = 0.0044 (FM) | $J25^h$ = 0.0050 (FM) |
| $j(X)^d$ (Å$^{-1}$) | $j$(O5): 0.0022x2 | $j$(O4): 0.0025x2 |
| ($\Delta h(X)^e$ Å, $l_n$'/$l_n^f$, CuXCu$^g$) | (0.259, 4.1, 101.8°) | (0.266, 3.9, 100.6°) |
| $Jn/J$max | $J25/J10$ = -0.04 | $J25/J10$ = -0.04 |
| Bond | Cu2-Cu3 | Cu1-Cu3 |
| d(Cu-Cu) (Å) | 5.424 | 5.269 |
| $J26^{(c)}$ (Å$^{-1}$) | $J26^h$ = 0.0022 (FM) ↔ -0.0055 (AFM) | $J26^h$ = 0.0034 (FM) ↔ -0.0285 (AFM) |
| $j(X)^d$ (Å$^{-1}$) | $j$(O9): 0.0025 (FM) | $j$(O9): 0.0034 |

| ($\Delta h$(X)[e] Å, $l_n'/l_n$[f], CuXCu[g]) | (0.297, 4.0, 101.3°) | (0.338, 3.6, 100.6°) |
|---|---|---|
| $J$n/$J$max | $J26/J10 = -0.02$ | $J26/J10 = -0.03$ |
| Bond | Cu1-Cu4 | Cu2-Cu4 |
| d(Cu-Cu) (Å) | 5.561 | 5.384 |
| $J27$[(c)] (Å$^{-1}$) | $J27$[h] = -0.0074 (AFM) | $J27$[h] = -0.0089 (AFM) |
| $j$(X)[d] (Å$^{-1}$) | $j$(O2): -0.0074 | $j$(O3): -0.0083 |
| ($\Delta h$(X)[e] Å, $l_n'/l_n$[f], CuXCu[g]) | (-0.480, 2.1, 139.1°) | (-0.485, 2.03, 138.6°) |
| $J$n/$J$max | $J27/J10 = 0.06$ | $J27/J10 = 0.06$ |
| Bond | Cu2-Cu3 | Cu1-Cu3 |
| d(Cu-Cu) (Å) | 5.606 | 5.398 |
| $J28$[(c)] (Å$^{-1}$) | $J28 = -0.0013$ (AFM) | $J28 = -0.0006$ (AFM) |
| $j$(X)[d] (Å$^{-1}$) | $j$(O1):- 0.0013 | $j$(O2): -0.0006 |
| ($\Delta h$(X)[e] Å, $l_n'/l_n$[f], CuXCu[g]) | (-0.110, 2.8, 121.6°) | (-0.047, 2.9, 117.3°) |
| $J$n/$J$max | $J28/J10 = 0.01$ | $J28/J10 = 0.005$ |
| Bond | Cu1-Cu2 | Cu2-Cu3 |
| d(Cu-Cu) (Å) | 5.618 | 5.451 |
| $J29$[(c)] (Å$^{-1}$) | [i]$J29$[h] = -0.0118 (AFM) | [i]$J29$[h] = **-0.0652** (AFM) |
| $j$(X)[d] (Å$^{-1}$) | $j$(O9): -0.0122 | $j$(O9): = -0.0652 |
| ($\Delta h$(X)[e] Å, $l_n'/l_n$[f], CuXCu[g]) | (-0.791, 2.1, 152.5°) | (-0.781, 1.97, 151.6°) |
| $J$n/$J$max | $J29/J10 = 0.10$ | $J29/J10 = 0.55$ |
| Bond | Cu2-Cu4 | Cu3-Cu4 |
| d(Cu-Cu) (Å) | 5.853 | 5.824 |
| $J30$[(c)] (Å$^{-1}$) | $J30$[h] = -0.0254 (AFM) | $J30$[h] = -0.0222 (AFM) |
| $j$(X)[d] (Å$^{-1}$) | $j$(Cu4): -0.0223 | $j$(Cu4): -0.0195 |
| ($\Delta h$(X)[e] Å, $l_n'/l_n$[f], CuXCu[g]) | (-0.382, 1.0, 172.7°) | (-0.330, 1.0, 170.6°) |
| $J$n/$J$max | $J30/J10 = 0.22$ | $J30/J10 = 0.19$ |
| Bond | Cu1-Cu4 | Cu2-Cu4 |
| d(Cu-Cu) (Å) | 5.863 | 5.880 |
| $J31$[(c)] (Å$^{-1}$) | $J31$[h] = -0.0155 (AFM) | $J31$[h] = -0.0152 (AFM) |
| $j$(X)[d] (Å$^{-1}$) | $j$(O2): -0.0133 | $j$(O3): -0.0124 |
| ($\Delta h$(X)[e] Å, $l_n'/l_n$[f], CuXCu[g]) | (-0.942, 2.1, 159.9°) | (-0.888, 2.1, 157.6°) |
| $J$n/$J$max | $J31/J10 = 0.13$ | $J31/J10 = 0.13$ |
| Bond | Cu1-Cu1 | Cu2-Cu2 |
| d(Cu-Cu) (Å) | 5.902 | 5.690 |
| $J32$[(c)] (Å$^{-1}$) | [i]$J32$[h] = 0.0030 (FM)→0.0626 FM | [i]$J32$[h] = 0.0022 (FM) |
| $j$(X)[d] (Å$^{-1}$) | $j$(O8): 0.0015x2 | $j$(O7): 0.0011x2 |
| ($\Delta h$(X)[e] Å, $l_n'/l_n$[f], CuXCu[g]) | (0.240, 4.5, 104.4°) | (0.128, 3,6 110°) |
| $J$n/$J$max | [i]$J32/J10 = -0.03 \rightarrow -0.54$ | $J32/J10 = -0.02$ |
| Bond | Cu2-Cu4 | Cu3-Cu4 |
| d(Cu-Cu) (Å) | 6.143 | 6.081 |
| $J33$[(c)] (Å$^{-1}$) | $J33 = -0.0178$ (AFM) | $J33 = -0.0017$ (AFM) |
| $j$(X)[d] (Å$^{-1}$) | $j$(O1): -0.0065 | $j$(O3): 0.0015 |
| ($\Delta h$(X)[e] Å, $l_n'/l_n$[f], CuXCu[g]) | (-0.599, 2.4, 145.5°) | (0.264, 4.7, 104.5°) |
| $j$(X)[d] (Å$^{-1}$) | $j$(O5): -0.0113 | $j$(O4): -0.0032 |
| ($\Delta h$(X)[e] Å, $l_n'/l_n$[f], CuXCu[g]) | (-0.937, 2.2, 160.2°) | (-0.301, 2.5, 133.5°) |
| $J$n/$J$max | $J33/J10 = 0.15$ | $J33/J10 = 0.01$ |
| Bond |  | Cu1-Cu4 |
| d(Cu-Cu) (Å) |  | 6.457 |
| $J33'$[(c)] (Å$^{-1}$) |  | $J33' = -0.0076$ (AFM) |
| $j$(X)[d] (Å$^{-1}$) |  | $j$(O2): -0.0076 |
| ($\Delta h$(X)[e] Å, $l_n'/l_n$[f], CuXCu[g]) |  | (-0.789, 2.5, 154.2°) |
| $J$n/$J$max |  | $J33'/J10 = 0.06$ |
| Bond | Cu3-Cu4 | Cu1-Cu4 |
| d(Cu-Cu) (Å) | 6.162 | 6.051 |
| $J34$[(c)] (Å$^{-1}$) | $J34 = -0.0038$ (AFM) | $J34 = -0.0090$ (AFM) |
| $j(X)^d$ (Å$^{-1}$) | $j$(O2): -0.0021 | $j$(O3): -0.0030 |
| ($\Delta h(X)^e$ Å, $l_n$'/$l_n^f$, CuXCu$^g$) | (-0.225, 2.8, 129.7°) | (-0.288, 2.6, 132.0°) |
| $j(X)^d$ (Å$^{-1}$) | $j$(O9): -0.0055 | $j$(O9): -0.0072 |
| ($\Delta h(X)^e$ Å, $l_n$'/$l_n^f$, CuXCu$^g$) | (-0.504, 2.4, 141.9°) | (-0.569, 2.2, 145.2°) |
| $j(X)^d$ (Å$^{-1}$) | $j$(O8): 0.0038 | $j$(O7): 0.0012 |
| ($\Delta h(X)^e$ Å, $l_n$'/$l_n^f$, CuXCu$^g$) | (0.072, 1.0, 128.9.°) | (0.023, 1.0, 129.6.°) |
| $J$n/$J$max | $J34/J10 = 0.15$ | $J34/J10 = 0.08$ |
| Bond | Cu1-Cu3 | Cu2-Cu1 |
| d(Cu-Cu) (Å) | 6.177 | 5.974 |
| $J35^{(c)}$ (Å$^{-1}$) | $J35 = -0.0185$ (AFM) | $J35 = -0.0224$ (AFM) |
| $j(X)^d$ (Å$^{-1}$) | $j$(O4): -0.0128 | $j$(O5): -0.0167 |
| ($\Delta h(X)^e$ Å, $l_n$'/$l_n^f$, CuXCu$^g$) | (-1.088, 2.2, 166.6°) | (-1.237, 2.1, 172.9°) |
| $j(X)^d$ (Å$^{-1}$) | $j$(O6): -0.0057 | $j$(O6): -0.0057 |
| ($\Delta h(X)^e$ Å, $l_n$'/$l_n^f$, CuXCu$^g$) | (-0.552, 2.6, 143.2°) | (-0.510, 2.5, 140.6°) |
| $J$n/$J$max | $J35/J10 = 0.14$ | $J35/J10 = 0.14$ |
| Bond | Cu3-Cu4 | Cu1-Cu4 |
| d(Cu-Cu) (Å) | 6.210 | 6.071 |
| $J36^{(c)}$ (Å$^{-1}$) | $^iJ36 = -0.0141$ (AFM) | $^iJ36 = -0.0656 \leftrightarrow -0.0160$ AFM |
| $j(X)^d$ (Å$^{-1}$) | $j$(O9): -0.0109 | $j$(O9): -0.0621 ↔ -0.0125 |
| ($\Delta h(X)^e$ Å, $l_n$'/$l_n^f$, CuXCu$^g$) | (-0.915, 2.2, 159.6°) | (-0.918, 1.99, 159.6°) |
| $j(X)^d$ (Å$^{-1}$) | $j$(O2): -0.0032 | $j$(O3): -0.0035 |
| ($\Delta h(X)^e$ Å, $l_n$'/$l_n^f$, CuXCu$^g$) | (-0.335, 2.8, 134.1°) | (-0.336, 2.6, 133.9°) |
| $J$n/$J$max | $J36/J10 = 0.12$ | $J36/J10 = ^i 0.55 \leftrightarrow 0.13$ |
| Bond | Cu1-Cu2 | Cu2-Cu3 |
| d(Cu-Cu) (Å) | 6.243 | 5.998 |
| $J37^{(c)}$ (Å$^{-1}$) | $^iJ37 = 0.0089$ (FM) ↔ -0.0159 (AFM) | $^iJ37 = -0.0620$ AFM |
| $j(X)^d$ (Å$^{-1}$) | $j$(O1): 0.0248 (FM) | $j$(O2): 0.0146 (FM) |
| ($\Delta h(X)^e$ Å, $l_n$'/$l_n^f$, CuXCu$^g$) | (0.456, 1.4, 117.5°) | (0.245, 1.4, 121.2°) |
| $j(X)^d$ (Å$^{-1}$) | $j$(O5): -0.0127 (AFM) | $j$(O4): -0.0726 (AFM) |
| ($\Delta h(X)^e$ Å, $l_n$'/$l_n^f$, CuXCu$^g$) | (-1.038, 2.10, 164.9°) | (-1.055, 1.96, 165.4°) |
| $j(X)^d$ (Å$^{-1}$) | $j$(O6): -0.0032 (AFM) | $j$(O6): -0.0040 (AFM) |
| ($\Delta h(X)^e$ Å, $l_n$'/$l_n^f$, CuXCu$^g$) | (-0.356, 2.8, 134.6°) | (-0.390, 2.7, 135.2°) |
| $J$n/$J$max | $J37/J10 = -0.08 \leftrightarrow 0.14$ | $J37/J10 = 0.52$ |
| Bond | Cu1-Cu2 | Cu2-Cu3 |
| d(Cu-Cu) (Å) | 6.245 | 6.007 |
| $^iJ38^{(c)}$ (Å$^{-1}$) | $^iJ38 = -0.0171$ (AFM) | $^iJ38 = -0.0198 \leftrightarrow -0.0819$ AFM |
| $j(X)^d$ (Å$^{-1}$) | $j$(O7): -0.0129 | $j$(O8): -0.0157 ↔ -0.0778 AFM |
| ($\Delta h(X)^e$ Å, $l_n$'/$l_n^f$, CuXCu$^g$) | (-1.097, 2.2, 167.2°) | (-1.126, 1.99, 168.3°) |
| $j(X)^d$ (Å$^{-1}$) | $j$(O9): -0.0042 | $j$(O9): -0.0041 |
| ($\Delta h(X)^e$ Å, $l_n$'/$l_n^f$, CuXCu$^g$) | (-0.443, 2.7, 138.6°) | (-0.394, 2.6, 135.6°) |
| $J$n/$J$max | $J38/J10 = 0.15$ | $J38/J10 = 0.17 \leftrightarrow 0.69$ |
| Bond | Cu3-Cu4 | Cu1-Cu4 |
| d(Cu-Cu) (Å) | 6.370 | 6.373 |
| $J39^{(c)}$ (Å$^{-1}$) | $J39 = -0.0083$ (AFM) | $J39 = -0.0066$ AFM |
| $j(X)^d$ (Å$^{-1}$) | $j$(O2): -0.0083 | $j$(O2): -0.0066 |
| ($\Delta h(X)^e$ Å, $l_n$'/$l_n^f$, CuXCu$^g$) | (-0.786, 2.3, 154.4°) | (-0.677, 2.5, 149.3°) |
| $J$n/$J$max | $J39/J10 = 0.07$ | $J39/J10 = 0.06$ |
| Bond | Cu4-Cu4 | Cu4-Cu4 |
| d(Cu-Cu) (Å) | 6.410 | 6.185 |
| $J40^{(c)}$ (Å$^{-1}$) | $J40^h = 0.0224$ (FM) | $J40^h = 0.0426$ (FM) |
| $j(X)^d$ (Å$^{-1}$) | $j$(O8): 0.0115x2 | $j$(O7): 0.0216x2 |
| ($\Delta h(X)^e$ Å, $l_n$'/$l_n^f$, CuXCu$^g$) | (0.233, 1.2, 125.8°) | (0.403, 1.3, 119.0°) |
| $J$n/$J$max | $J40/J10 = -0.19$ | $J40/J10 = -0.36$ |
| Bond | Cu1-Cu4 | Cu2-Cu4 |
| d(Cu-Cu) (Å) | 6.426 | 6.276 |
| $J41^{(c)}$ (Å$^{-1}$) | $J41^h$ = -0.0280 (AFM) | $J41^h$ = -0.0321 (AFM) |
| $j(X)^d$ (Å$^{-1}$) | $j$(O7): -0.0284 | $j$(O8): -0.0344 |
| $(\Delta h(X)^e$ Å, $l_n'/l_n^f$, CuXCu$^g$) | (-0.583, 1.1, 151.4°) | (-0.677, 1.1, 154.0°) |
| $J$n/$J$max | $J41/J10$ = 0.24 | $J41/J10$ = 0.27 |
| Bond | Cu3-Cu4 | Cu1-Cu4 |
| d(Cu-Cu) (Å) | 6.482 | 6.312 |
| $J42^{(c)}$ (Å$^{-1}$) | $J42^h$ = -0.0224 (AFM) | $J42^h$ = -0.0219 (AFM) |
| $j(X)^d$ (Å$^{-1}$) | $j$(Cu1): -0.0187 | $j$(Cu1): -0.0206 |
| $(\Delta h(X)^e$ Å, $l_n'/l_n^f$, CuXCu$^g$) | (-0.384, 1.24, 173.4°) | (-0.403, 1.2, 173.9°) |
| $J$n/$J$max | $J42/J10$ = 0.19 | $J42/J10$ = 0.18 |
| Bond | Cu2-Cu4 | Cu3-Cu4 |
| d(Cu-Cu) (Å) | 6.505 | 6.328 |
| $J43^{(c)}$ (Å$^{-1}$) | $^i J43$ = 0.0227 (FM) | $^i J43$ = -0.0025 (AFM) |
| $j(X)^d$ (Å$^{-1}$) | $j$(As1): 0.0209 | $^i$- |
| $(\Delta h(X)^e$ Å, $l_n'/l_n^f$, CuXCu$^g$) | (0.441, 1.1, 153.1°) | $^i$- |
| $j(X)^d$ (Å$^{-1}$) | $j$(O2): -0.0031 | $j$(O3): -0.0047 |
| $(\Delta h(X)^e$ Å, $l_n'/l_n^f$, CuXCu$^g$) | (-0.376, 2.9, 136.6°) | (-0.486, 2.6, 141.2°) |
| $j(X)^d$ (Å$^{-1}$) | $j$(O7): 0.0049 | $j$(O8): 0.0022 |
| $(\Delta h(X)^e$ Å, $l_n'/l_n^f$, CuXCu$^g$) | (0.103, 1.1, 130.3°) | (0.043, 1.1, 130.8) |
| $J$n/$J$max | $^i J43/J10$ = -0.20 | $^i J43/J10$ = 0.02 |
| Bond | Cu1-Cu3 | Cu1-Cu2 |
| d(Cu-Cu) (Å) | 6.627 | 6,420 |
| $J44^{(c)}$ (Å$^{-1}$) | $J44$ = 0.0194 (FM) | $J44$ = 0.0137 (FM) |
| $j(X)^d$ (Å$^{-1}$) | $j$(Cu2): 0.0158 (FM) | $j$(Cu3): 0.0184 (FM) |
| $(\Delta h(X)^e$ Å, $l_n'/l_n^f$, CuXCu$^g$) | (0.326, 1.4, 148.9°) | (0.361, 1.4, 146.9°) |
| $j(X)^d$ (Å$^{-1}$) | $j$(O5): 0.0084 (FM) | $j$(O5): 0.0010 (FM) |
| $(\Delta h(X)^e$ Å, $l_n'/l_n^f$, CuXCu$^g$) | (0.175, 1.4, 128.1°) | (0.182, 4.6, 109.0°) |
| $j(X)^d$ (Å$^{-1}$) | $j$(O3): -0.0051 | $j$(O1): -0.0057 |
| $(\Delta h(X)^e$ Å, $l_n'/l_n^f$, CuXCu$^g$) | (-0.621, 2.8, 147.0°) | (-0.621, 2.7, 146.6°) |
| $J$n/$J$max | $J44/J10$ = -0.17 | $J44/J10$ = -0.12 |
| Bond | Cu1-Cu2 | Cu2-Cu3 |
| d(Cu-Cu) (Å) | 6.696 | 6.677 |
| $J45'^{(c)}$ (Å$^{-1}$) | $J45'^h$ = -0.0010 (AFM) | $J45'^h$ = -0.0004 |
| $j(X)^d$ (Å$^{-1}$) | $j$(O7): -0.0010 | $j$(O8): -0.0004 |
| $(\Delta h(X)^e$ Å, $l_n'/l_n^f$, CuXCu$^g$) | (-0.152, 3.3, 127.5°) | (-0.065, 3.4, 124.3°) |
| $J$n/$J$max | $J45'/J10$ = 0.009 | $J45'/J10$ = 0.003 |
| Bond | Cu2-Cu4 | Cu3-Cu4 |
| d(Cu-Cu) (Å) | 7.050 | 6.881 |
| $J46^{(c)}$ (Å$^{-1}$) | $J46$ = 0.0089 (FM) ↔ 0.0299 (FM) | $J46$ = 0.0324 (FM) |
| $j(X)^d$ (Å$^{-1}$) | $j$(Cu2): 0.0210 (FM) | $j$(Cu3): 0.0216 (FM) |
| $(\Delta h(X)^e$ Å, $l_n'/l_n^f$, CuXCu$^g$) | (0.472, 1.6, 145.6°) | (0.470, 1.5, 145.1°) |
| $j(X)^d$ (Å$^{-1}$) | $j$(O7): 0.0121 (FM) | $j$(O8): 0.0164 (FM) |
| $(\Delta h(X)^e$ Å, $l_n'/l_n^f$, CuXCu$^g$) | (0.298, 1.1, 128.4°) | (0.385, 1.2, 125.0°) |
| $j(X)^d$ (Å$^{-1}$) | $j$(O3): -0.0052 (AFM) | $j$(O1): -0.0056 (AFM) |
| $(\Delta h(X)^e$ Å, $l_n'/l_n^f$, CuXCu$^g$) | (-0.750, 2.9, 153.0°) | (-0.747, 2.8, 152.7°) |
| $J$n/$J$max | $J46/J10$ = -0.08 ↔ -0.26 | $J46/J10$ = -0.27 |
| Bond | Cu3-Cu4 | Cu1-Cu4 |
| d(Cu-Cu) (Å) | 7.197 | 7.026 |
| $J47^{(c)}$ (Å$^{-1}$) | $J47^h$ = 0.0068 (FM) | $J47^h$ = 0.0068 (FM) |
| $j(X)^d$ (Å$^{-1}$) | $j$(O3): -0.0066 (AFM) | $j$(O1): -0.0074 (AFM) |
| $(\Delta h(X)^e$ Å, $l_n'/l_n^f$, CuXCu$^g$) | (-0.981, 2.9, 162.8°) | (-1.004, 2.8, 163.7°) |
| $j(X)^d$ (Å$^{-1}$) | $j$(O9): 0.0116 (FM) | $j$(O9): 0.0120 (FM) |
| $(\Delta h(X)^e$ Å, $l_n'/l_n^f$, CuXCu$^g$) | (0.298, 1.1, 130.0°) | (0.292, 1.2, 128.3°) |
| $J$n/$J$max | $J47/J10$ = 0.06 | $J47/J10$ = 0.06 |

| Bond | Cu1-Cu3 | Cu1-Cu2- |
|---|---|---|
| d(Cu-Cu) (Å) | 7.213 | 6.924 |
| $J48^{(c)}$ (Å$^{-1}$) | $J48^h$ = -0.0484 (AFM) | $J48^h$= -0.0552 (AFM) |
| $j(X)^d$ (Å$^{-1}$) | $j$(O5): -0.0492 | $j$(O4): -0.0552 |
| ($\Delta h(X)^e$ Å, $l_n$'/$l_n^f$, CuXCu$^g$) | (-1.243, 1.3, 174.9°) | (-1.275, 1.3, 175.8°) |
| $Jn/J$max | $J48/J10$ = 0.42 | $J48/J10$ = 0.46 |
| Bond | Cu2-Cu3 | Cu3-Cu1 |
| d(Cu-Cu) (Å) | 7.316 | 7.263 |
| $J49^{(c)}$ (Å$^{-1}$) | $J49^h$ = 0.0167 (FM) | $J49^h$ = 0.0208 (FM) |
| $j(X)^d$ (Å$^{-1}$) | $j$(Cu1): 0.0201(FM) | $j$(Cu2): 0.0247(FM) |
| ($\Delta h(X)^e$ Å, $l_n$'/$l_n^f$, CuXCu$^g$) | (0.467, 1.7, 146.3°) | (0.559, 1.8, 143.0°) |
| $j(X)^d$ (Å$^{-1}$) | $j$(O3): -0.0038 (AFM) | $j$(O1): -0.0041 (AFM) |
| ($\Delta h(X)^e$ Å, $l_n$'/$l_n^f$, CuXCu$^g$) | (-0.648, 3.2, 149.0°) | (-0.674, 3.1, 150.0°) |
| $Jn/J$max | $J49/J10$ = -0.14 | $J49/J10$ = -0.14 |
| Bond | Cu1-Cu3 | Cu2-Cu1 |
| d(Cu-Cu) (Å) | 7.432 | 7.322 |
| $J50^{(c)}$ (Å$^{-1}$) | **i**$J50^h$ = 0.0010 (FM) | **i**$J50^h$ = 0.0197 ↔ 0.0017(FM) |
| $j(X)^d$ (Å$^{-1}$) | $j$(As1): 0.0044 (FM) | $j$(P1): 0.0117 (FM) |
| ($\Delta h(X)^e$ Å, $l_n$'/$l_n^f$, CuXCu$^g$) | (0.120, 1.0, 166.3°) | (0.310, 1.1, 165.0°) |
| $j(X)^d$ (Å$^{-1}$) | $j$(O9): -0.0034 (AFM) | $j$(O9): -0.0100 (AFM) |
| ($\Delta h(X)^e$ Å, $l_n$'/$l_n^f$, CuXCu$^g$) | (-0.093, 1.1, 141.2°) | (-0.267, 1.1, 145.3°) |
| $j(X)^d$ (Å$^{-1}$) | - | **i**$j$(O3): 0.0180 (FM)↔0 |
| ($\Delta h(X)^e$ Å, $l_n$'/$l_n^f$, CuXCu$^g$) | - | (0.482, 1.1, 125.5°) |
| $Jn/J$max | $J50/J10$ = -0.009 | $J50/J10$ = -0.17 |
| Bond | Cu1-Cu4 | Cu2-Cu4 |
| d(Cu-Cu) (Å) | 7.465 | 7.389 |
| $J51^{(c)}$ (Å$^{-1}$) | $J51^h$ = -0.0028 (AFM) | $J51^h$ = -0.0019 (AFM) |
| $j(X)^d$ (Å$^{-1}$) | $j$(O8): -0.0039 (AFM) | $j$(O7): -0.0032 (AFM) |
| ($\Delta h(X)^e$ Å, $l_n$'/$l_n^f$, CuXCu$^g$) | (-0.681, 3.1, 151.2°) | (-0.560, 3.2, 146.2°) |
| $Jn/J$max | $J51/J10$ = -0.02 | $J51/J10$ = -0.02 |
| Bond | Cu1-Cu4 | Cu2-Cu4 |
| d(Cu-Cu) (Å) | 7.467 | 7.237 |
| $J52^{(c)}$ (Å$^{-1}$) | $J^i52^h$ = -0.0338 (AFM) | $J52^h$ = -0.0322 (AFM) |
| $j(X)^d$ (Å$^{-1}$) | $j$(O7): -0.0332 | $j$(O8): -0.0302 |
| ($\Delta h(X)^e$ Å, $l_n$'/$l_n^f$, CuXCu$^g$) | (-0.880, 1.4, 163.7°) | (-0.758, 1.4, 159.4°) |
| $Jn/J$max | $J52/J10$ = 0.29 | $J40/J10$ = 0.27 |
| Bond | Cu1-Cu4 | Cu2-Cu4 |
| d(Cu-Cu) (Å) | 7.517 | 7.284 |
| $J53^{(c)}$ (Å$^{-1}$) | $J53^h$ = 0.0185 (FM) | $J5^h3$ = 0.107 (FM) |
| $j(X)^d$ (Å$^{-1}$) | $j$(Cu3): 0.0016 | $j$(Cu1): -0.0025 |
| ($\Delta h(X)^e$ Å, $l_n$'/$l_n^f$, CuXCu$^g$) | (0.042, 1.5, 160.8°) | (-0.063, 1.4, 163.8°) |
| $j(X)^d$ (Å$^{-1}$) | $j$(O7): 0.0170 | $j$(O8): 0.0142 |
| ($\Delta h(X)^e$ Å, $l_n$'/$l_n^f$, CuXCu$^g$) | (0.479, 1.1, 126.8°) | (0.376, 1.1, 128.0°) |
| $Jn/J$max | $J53/J10$ = -0.16 | $J53/J10$ = -0.09 |
| Bond | Cu2-Cu4 | Cu3-Cu4 |
| d(Cu-Cu) (Å) | 7.596 | 7.371 |
| $J54^{(c)}$ (Å$^{-1}$) | $J54^h$ = 0.0263 (FM) ↔ 0.0068 (FM) | $J5^h4$ = 0.0244 (FM) |
| $j(X)^d$ (Å$^{-1}$) | $j$(Cu1): 0.0195 | $j$(Cu2): 0.0175 |
| ($\Delta h(X)^e$ Å, $l_n$'/$l_n^f$, CuXCu$^g$) | (0.474, 1.8, 146.8°) | (0.369, 1.7, 149.5°) |
| $j(X)^d$ (Å$^{-1}$) | $j$(O9): 0.0050 | $j$(O9): 0.0066 |
| ($\Delta h(X)^e$ Å, $l_n$'/$l_n^f$, CuXCu$^g$) | (0.142, 1.1, 135.7°) | (0.180, 1.0, 133.6°) |
| $Jn/J$max | $J54/J10$ = -0.23 ↔ -0.06 | $J54/J10$ = -0.21 |
| Bond | Cu1-Cu2 | Cu2-Cu3 |
| d(Cu-Cu) (Å) | 7.604 | 7.428 |
| $J55^{(c)}$ (Å$^{-1}$) | $J55^h$ = 0.0171 (FM) ↔ 0.0344 (FM) | $J55^h$ = 0.0309 FM |
| $j(X)^d$ (Å$^{-1}$) | $j$(O2): 0.0162 | $j$(O3): 0.0146 |
| ($\Delta h(X)^e$ Å, $l_n'/l_n^f$, CuXCu$^g$) | (0.458, 1.2, 127.5°) | (0.388, 1.3, 127.9°) |
| $j(X)^d$ (Å$^{-1}$) | $j$(O7): 0.0174 | $j$(O8): 0.0158 |
| ($\Delta h(X)^e$ Å, $l_n'/l_n^f$, CuXCu$^g$) | (0.491, 1.2, 126.7°) | (0.429, 1.2, 127.3°) |
| $J$n/$J$max | $J55/J10$ = -0.15 ↔ -030 | $J55/J10$ = -0.26 |
| Bond | Cu2-Cu3 | Cu1-Cu3 |
| d(Cu-Cu) (Å) | 7.618 | 7.548 |
| $J56^{(c)}$ (Å$^{-1}$) | $J56^h$ = -0.0178 (AFM) | $J56^h$ = -0.0211 (AFM) |
| $j(X)^d$ (Å$^{-1}$) | $j$(O2): -0.0176 | $j$(O3): -0.0206 |
| ($\Delta h(X)^e$ Å, $l_n'/l_n^f$, CuXCu$^g$) | (-0.500, 1.2, 153.2°) | (-0.584, 1.1, 155.5°) |
| $J$n/$J$max | $J56/J10$ = 0.15 | $J56/J10$ = 0.18 |
| Bond | Cu4-Cu4 | Cu4-Cu4 |
| d(Cu-Cu) (Å) | 7.787 | 7.576 |
| $J57^{(c)}$ (Å$^{-1}$) | $J57^h$ = 0.0285 (FM) | $J57^h$ = 0.0281 (FM) |
| $j(X)^d$ (Å$^{-1}$) | $j$(O9): 0.0139x2 | $j$(O9): 0.0136x2 |
| ($\Delta h(X)^e$ Å, $l_n'/l_n^f$, CuXCu$^g$) | (0.419, 1.1, 129.8°) | (0.390, 1.0, 129.4°) |
| $J$n/$J$max | $J57/J10$ = -0.25 | $J57/J10$ = -0.24 |
| Bond | Cu2-Cu2 | Cu3-Cu3 |
| d(Cu-Cu) (Å) | 8.062 | 7.992 |
| $J58^{(c)}$ (Å$^{-1}$) | $J58^h$= 0.0096 (FM) | $J58^h$= 0.0016 (FM) |
| $j(X)^d$ (Å$^{-1}$) | $j$(O6): 0.0073x2 | $j$(O6): 0.0035x2 |
| ($\Delta h(X)^e$ Å, $l_n'/l_n^f$, CuXCu$^g$) | (0.234, 1.2, 135.5°) | (0.111, 1.1, 138.5°) |
| $j(X)^d$ (Å$^{-1}$) | $j$(O5): -0.0025x2 | $j$(O4): -0.0027x2 |
| ($\Delta h(X)^e$ Å, $l_n'/l_n^f$, CuXCu$^g$) | (-0.550, 3.3, 147.3°) | (-0.568, 3.3, 148.3°) |
| $J$n/$J$max | $J58/J10$ = -0.08 | $J58/J10$ = -0.01 |
| Bond | Cu1-Cu1 | Cu2-Cu2 |
| d(Cu-Cu) (Å) | 6.415 | 6.272 |
| $J_a^{1-1\,(c)}$ (Å$^{-1}$) | $J_a^{1-1h}$ = -0.0205 (AFM) | $J_c^{2-2h}$ = -0.0230(AFM) |
| $j(X)^d$ (Å$^{-1}$) | $j$(O8): -0.0117 | $j$(O7): -0.0123 |
| ($\Delta h(X)^e$ Å, $l_n'/l_n^f$, CuXCu$^g$) | (-1.108, 2.3, 167.7°) | (-1.084, 2.2, 166.6°) |
| $j(X)^d$ (Å$^{-1}$) | $j$(O9): -0.0088 | $j$(O9): -0.0107 |
| ($\Delta h(X)^e$ Å, $l_n'/l_n^f$, CuXCu$^g$) | (-0.883, 2.4, 158.0°) | (-0.975, 2.3, 161.8°) |
| $J$n/$J$max | $J_a^{1-1}/J10$ = 0.18 | $J_c^{2-2}/J10$ = 0.19 |
| Bond | Cu2-Cu2 | Cu3-Cu3 |
| d(Cu-Cu) (Å) | 6.415 | 6.272 |
| $J_a^{2-2-\,(c)}$ (Å$^{-1}$) | $J_a^{2-2\,h}$ = -0.0210 (AFM) | $J_c^{3-3h}$ = -0.0255 |
| $j(X)^d$ (Å$^{-1}$) | $j$(Cu3): -0.0160 | $j$(Cu1): -0.0210 |
| ($\Delta h(X)^e$ Å, $l_n'/l_n^f$, CuXCu$^g$) | (-0.328, 1.1, 171.4°) | (-0.413, 1.1, 174.2°) |
| $j(X)^d$ (Å$^{-1}$) | $j$(O1): -0.0042 | $j$(O2): -0.0036 |
| ($\Delta h(X)^e$ Å, $l_n'/l_n^f$, CuXCu$^g$) | (-0.468, 2.7, 140.4°) | (-0.395, 2.8, 136.6°) |
| $J$n/$J$max | $J_a^{2-2}/J10$ = 0.18 | $J_c^{3-3}/J10$ = 0.21 |
| Bond | Cu3-Cu3 | Cu1-Cu1 |
| d(Cu-Cu) (Å) | 6.415 | 6.272 |
| $J_a^{3-3-\,(c)}$ (Å$^{-1}$) | $J_a^{3-3}$ = -0.0215 (AFM) | $J_c^{1-1}$ ==0.0259 |
| $j(X)^d$ (Å$^{-1}$) | $j$(Cu2): -0.0160 | $j$(Cu3): -0.0210 |
| ($\Delta h(X)^e$ Å, $l_n'/l_n^f$, CuXCu$^g$) | (-0.328, 1.1, 171.4°) | (-0.413, 1.1, 174.2°) |
| $j(X)^d$ (Å$^{-1}$) | $j$(O3): -0.0036 | $j$(O1): -0.0029 |
| ($\Delta h(X)^e$ Å, $l_n'/l_n^f$, CuXCu$^g$) | (-0.426, 2.9, 137.8°) | (-0.341, 3.0, 133.4°) |
| $j(X)^d$ (Å$^{-1}$) | $j$(O1): -0.0019 | $j$(O2): -0.0020 |
| ($\Delta h(X)^e$ Å, $l_n'/l_n^f$, CuXCu$^g$) | (-0.240, 3.1, 130.2°) | (-0.240, 3.0, 129.7°) |
| $J$n/$J$max | $J_a^{3-3}/J10$ = 0.19 | $J_c^{1-1}/J10$ = 0.22 |
| Bond | Cu4-Cu4 | Cu4-Cu4 |
| d(Cu-Cu) (Å) | 6.415 | 6.272 |
| $J_a^{4-4\ (c)}$ (Å$^{-1}$) | $J_a^{4-4\ h}$ = -0.0016 (AFM) | $J_c^{4-4\ h}$ = -0.0013 (AFM) |
| $j(X)^d$ (Å$^{-1}$) | $j$(O5): -0.0019 | $j$(O4): -0.0019 |
| ($\Delta h(X)^e$ Å, $l_n'/l_n^f$, CuXCu$^g$) | (-0.239, 3.0, 130.4°) | (-0.226, 3.0, 129.5°) |
| $J$n/$J$max | $J_a^{4-4}/J10$ = 0.01 | $J_a^{4-4}/J10$ = 0.01 |
| Bond | Cu1-Cu1 | Cu2-Cu2 |
| d(Cu-Cu) (Å) | 7.655 | 7.528 |
| $J_b^{1-1\ (c)}$ (Å$^{-1}$) | $J_b^{1-1}$ = 0.0209 (FM) | $J_a^{2-2}$ = 0.0239 (FM) |
| $j(X)^d$ (Å$^{-1}$) | $j$(O7): 0.0144 | $j$(O8): 0.0198 |
| ($\Delta h(X)^e$ Å, $l_n'/l_n^f$, CuXCu$^g$) | (0.350, 1.89, 127.3°) | (0.453, 1.96, 123.5°) |
| $j(X)^d$ (Å$^{-1}$) | $j$(O1): 0.0065 | $j$(O2): 0.0041 |
| ($\Delta h(X)^e$ Å, $l_n'/l_n^f$, CuXCu$^g$) | (0.163, 1.75, 132.9°) | (0.093, 1.96, 132.9°) |
| $J$n/$J$max | $J_b^{1-1}/J10$ = -0.18 | $J_a^{2-2}/J10$ = -0.20 |
| Bond | Cu2-Cu2 | Cu3-Cu3 |
| d(Cu-Cu) (Å) | 7.655 | 7.528 |
| $J_b^{2-2\ (c)}$ (Å$^{-1}$) | $J_b^{2-2\ h}$ = -0.0095 (AFM) | $J_a^{3-3h}$ = -0.0063 (AFM) |
| $j(X)^d$ (Å$^{-1}$) | $j$(O6): -0.0111 | $j$(O6): -0.0080 |
| ($\Delta h(X)^e$ Å, $l_n'/l_n^f$, CuXCu$^g$) | (-0.324, 1.0, 148.6°) | (-0.225, 1.1, 145.3°) |
| $J$n/$J$max | $J_b^{2-2}/J10$ = 0.08 | $J_a^{3-3}/J10$ = 0.05 |
| Bond | Cu3-Cu3 | Cu1-Cu1 |
| d(Cu-Cu) (Å) | 7.655 | 7.528 |
| $J_b^{3-3\ (c)}$ (Å$^{-1}$) | $J_b^{3-3}$ = 0.0004 (FM) | $J_a^{1-1}$ = -0.0042 (AFM) |
| $j(X)^d$ (Å$^{-1}$) | $j$(O2): -0.0058 | $j$(O3): -0.0076 |
| ($\Delta h(X)^e$ Å, $l_n'/l_n^f$, CuXCu$^g$) | (-0.164, 1.3, 143.7°) | (-0.213, 1.2, 144.8°) |
| $j(X)^d$ (Å$^{-1}$) | $j$(O9): 0.0054 | $j$(O9): 0.0019 |
| ($\Delta h(X)^e$ Å, $l_n'/l_n^f$, CuXCu$^g$) | (0.157, 1.2, 135.5°) | (0.053, 1.2, 137.4°) |
| $j(X)^d$ (Å$^{-1}$) | $j$(O1): 0.0008 | $j$(O5): 0.0007 |
| ($\Delta h(X)^e$ Å, $l_n'/l_n^f$, CuXCu$^g$) | (0.332, 7.4, 103.4°) | (0.262, 6.4, 107.1°) |
| $J$n/$J$max | $J_b^{3-3}/J10$ = -0.003 | $J_a^{1-1}/J10$ = 0.03 |
| Bond | Cu4-Cu4 | Cu4-Cu4 |
| d(Cu-Cu) (Å) | 7.655 | 7.528 |
| $J_b^{4-4\ (c)}$ (Å$^{-1}$) | $J_b^{4-4h}$ = 0.0146 (FM) | $J_a^{4-4h}$ = 0.0126 (FM) |
| $j(X)^d$ (Å$^{-1}$) | $j$(O7): 0.0143 | $j$(O8): 0.0125 (FM) |
| ($\Delta h(X)^e$ Å, $l_n'/l_n^f$, CuXCu$^g$) | (0.419, 1.1, 129.1°) | (0.353, 1.0, 130.1°) |
| $J$n/$J$max | $J_b^{4-4}/J10$ = -0.13 | $J_a^{4-4}/J10$ = -0.11 |
| Bond | Cu1-Cu1 | Cu2-Cu2 |
| d(Cu-Cu) (Å) | 8.224 | 8.090 |
| $J_c^{1-1\ (c)}$ (Å$^{-1}$) | $J_c^{1-1h}$ = -0.0178 (AFM) | $J_b^{2-2h}$ = -0.0178 (AFM) |
| $j(X)^d$ (Å$^{-1}$) | $j$(O5): -0.0147 | $j$(O5): -0.0126 |
| ($\Delta h(X)^e$ Å, $l_n'/l_n^f$, CuXCu$^g$) | (-0.497, 1.0, 155.2°) | (-0.412, 1.1, 152.5°) |
| $J$n/$J$max | $J_c^{1-1}/J10$ = 0.15 | $J_b^{2-2}/J10$ = 0.15 |
| Bond | Cu2-Cu2 | Cu3-Cu3 |
| d(Cu-Cu) (Å) | 8.224 | 8.090 |
| $J_c^{2-2\ (c)}$ (Å$^{-1}$) | $J_c^{2-2h}$ = -0.0215 (AFM) | $J_b^{3-3h}$ = -0.0347 (AFM) |
| $j(X)^d$ (Å$^{-1}$) | $j$(O1): -0.0199 | $j$(O8): -0.0345 |
| ($\Delta h(X)^e$ Å, $l_n'/l_n^f$, CuXCu$^g$) | (-0.673, 1.0, 159.9°) | (-1.107, 1.2, 171.6°) |
| $J$n/$J$max | $J_c^{2-2}/J10$ = 0.19 | $J_b^{3-3}/J10$ = 0.29 |
| Bond | Cu3-Cu3 | Cu1-Cu1 |
| d(Cu-Cu) (Å) | 8.224 | 8.090 |
| $J_c^{3-3\ (c)}$ (Å$^{-1}$) | $J_c^{3-3}$ = -0.0028 (AFM) | $J_b^{1-1}$ = -0.0121 (AFM) |
| $j(X)^d$ (Å$^{-1}$) | $j$(O4): -0.0016 | $j$(O2): -0.0214 |
| ($\Delta h(X)^e$ Å, $l_n'/l_n^f$, CuXCu$^g$) | (-0.419, 3.9, 141.2°) | (-0.699, 1.1, 160.3°) |
| $j(X)^d$ (Å$^{-1}$) | $j$(O9): -0.0012 | $j$(O5): -0.0032 |
| ($\Delta h(X)^e$ Å, $l_n'/l_n^f$, CuXCu$^g$) | (-0.307, 3.9, 137.6°) | (-0.727, 3.4, 153.6°) |
| $j(X)^d$ (Å$^{-1}$) | - | $j$(O6): 0.0124 |
| ($\Delta h(X)^e$ Å, $l_n'/l_n^f$, CuXCu$^g$) | - | (0.406, 1.0, 131.9°) |

| $Jn/J$max | $J_c^{3-3}/J10 = 0.02$ | $J_b^{1-1}/J10 = 0.10$ |
|---|---|---|
| Bond | Cu4-Cu4 | Cu4-Cu4 |
| d(Cu-Cu) (Å) | 8.224 | 8.090 |
| $J_c^{4-4\ (c)}$ (Å$^{-1}$) | $J_c^{4-4h}$ = -0.0176 (AFM) | $J_b^{4-4h}$ = -0.0277 (AFM) |
| $j(X)^d$ (Å$^{-1}$) | $j$(O9): -0.0141 | $j$(O7): -0.0229 |
| ($\Delta h(X)^e$ Å, $l_n'/l_n^f$, Cu$X$Cu$^g$) | (-0.476, 1.0, 154.7°) | (-0.746, 1.1, 161.5°) |
| $Jn/J$max | $J_c^{4-4}/J10 = 0.15$ | $J_c^{4-4}/J10 = 0.23$ |

[a]XDS: X-ray diffraction from a single crystal.
[b]The refinement converged to the residual factor ($R$) values.
[c]$Jn$ in Å$^{-1}$: the magnetic couplings ($Jn < 0$, AFM; $Jn > 0$, FM). To translate the $Jn$ value in per angstrom (Å−1) into energy units more conventional for experimenters—millielectronvolt (meV)—one can use the scaling factor K = 74 ($Jn$ (meV) = 74 $Jn$ (A$^{-1}$))
[d]$j$(X): contributions of the intermediate ion X to the AFM ($j$(X) <0) and FM ($j$(X)>0) components of the coupling $Jn$.
[e]$\Delta h$(X): the degree of overlapping of the local space between magnetic ions by the intermediate ion X.
[f]$l_n'/l_n$: the asymmetry of the position of the intermediate ion X relative to the middle of the Cu$_i$–Cu$_j$ bond line.
[g]Cu$_i$XCu$_j$: bonding angle.
[h]Small j(X) contributions are not shown.
[i]The difference between arsenate and phosphate.